\documentclass[a4paper,11pt]{article}
\pdfoutput=1
\usepackage{jheppub}

\usepackage{multirow}
\newcommand{\superiso}{Mahmoudi:2007vz,Mahmoudi:2008tp}
\newcommand{\higgsbounds}{Bechtle:2008jh,Bechtle:2011sb}

\newcommand{\hzero}{h^0} %light neutral Higgs
\newcommand{\Hzero}{H^0} %heavy neutral Higgs
\newcommand{\Azero}{A^0} % CP-odd neutral Higgs
\title{Signal background interference effects in heavy scalar production and decay to a top--anti-top pair}

\author{B.~Hespel, F.~Maltoni, E.~Vryonidou}
\affiliation{Centre for Cosmology, Particle Physics and Phenomenology (CP3),\\
 Universit\'e catholique de Louvain, B-1348 Louvain-la-Neuve, Belgium}
\abstract{We analyse the production of a top quark pair through a heavy scalar at the LHC. We first review the main features of the signal as well as the interference with the top--anti-top background at leading order in QCD. We then study higher order QCD effects. While the background and the signal can be obtained at NNLO and NLO in QCD respectively, that is not the case for their interference, which is currently known only at LO. In order to improve the accuracy of the prediction for the interference term, we consider the effects of extra QCD radiation, i.e. the $2 \to 3$ (loop-induced) processes and obtain an estimate of the NLO corrections. As a result, we find that the contribution of the interference is important both at the total cross-section level and, most importantly, for the line-shape of the heavy scalar.  In particular for resonances with widths larger than a couple of percent of the resonance mass, the interference term distorts the invariant mass distribution and generically leads to a non-trivial peak-dip structure. We study this process in a simplified model involving an additional scalar or pseudoscalar resonance as well as in the Two-Higgs-Doublet-Model for a set of representative benchmarks. We present the constraints on simplified models featuring an extra scalar as set by the LHC searches for top--anti-top resonances, and the implications of the 750 GeV diphoton excess recently reported by CMS and ATLAS for the top pair production assuming a scalar or a pseudoscalar resonance. }

\preprint{CP3-16-30, MCnet-16-19}
\begin{document}
\maketitle

\section{Introduction}
\label{sec:intro}
The top quark might have a special role in electroweak symmetry breaking (EWSB), as it is the only fermion with a coupling to the Higgs of order one and therefore with a mass of order of the Higgs field vacuum expectation value. This unique feature has been exploited in a wide range of beyond the Standard Model (BSM) scenarios, from being a window into strongly interacting scenarios to triggering EWSB in supersymmetric theories.   After the top discovery at the Tevatron more than 20 years ago, a plethora of searches have been performed to measure its properties and to look for hints of new physics both at the Tevatron and the LHC~\cite{Agashe:2014kda}. 

The main production channel at hadron colliders is in $t\bar{t}$ pairs via strong interactions, with the EW single production following with roughly a third of the cross-section.  New physics effects in top pair production can generally be classified into two categories: resonant and non-resonant. Non-resonant effects are conveniently described within effective field theory, i.e. including the effect of dimension-6 operators  \cite{AguilarSaavedra:2008zc, Zhang:2010dr, Degrande:2010kt}. Resonant effects arise in physics models which predict new particles that couple to the top, either via an $s$-channel or a $t$-channel, leading to top quark pairs  possibly in association with other visible or invisible final states. The search for $s$-channel resonances above the $2m_t$ threshold but within the experimental reach, is particularly promising~\cite{Barger:2006hm}. These can be spin-0, 1 or 2, colour octet or singlet depending on the model \cite{Frederix:2007gi}. These resonances often arise only in top pair production if their couplings to light particles are suppressed.

Experimental searches for heavy scalar particles decaying into top quark pairs have been performed by both CMS \cite{Chatrchyan:2013lca}  and ATLAS \cite{Aad:2015fna} in Run I of the LHC. These searches are interpreted in terms of upper bounds on production cross-section times branching ratio, assume a narrow width resonance, and generally ignore the interference of the signal with the SM background. Nevertheless, it has been shown that the interference should be taken into account, as the heavy state does not necessarily show up as a resonance bump in the top pair invariant mass distribution but most likely leads to a peak-dip structure \cite{Gaemers:1984sj,Dicus:1994bm}.  Similar effects have been discussed recently in \cite{Djouadi:2016ack} in the light of the excess reported at 750 GeV by ATLAS \cite{atlasdiphoton} and CMS \cite{CMS:2015dxe}.

In order to extract maximal information on the new physics in the presence of an excess or to constrain  BSM scenarios in the absence of one, accurate predictions are not only needed for  signal and background,  but also for their interference. The QCD background is known at NNLO in QCD \cite{Czakon:2013goa} and NLO in electroweak \cite{Bernreuther:2006vg} (see also~\cite{Pagani:2016caq} for a detailed study at LHC 13 TeV).  Recent work in the direction of promoting signal and interference predictions beyond LO has been presented in \cite{Bernreuther:2015fts}.  While all ingredients are available for the computation of the signal at NLO, this is often computed at LO or in some approximation such as the one in \cite{Bernreuther:2015fts}. NLO $K$-factors computed for the scalar production cross-section, $\sigma(p p \to \Phi)$, are often applied to the signal, especially in studies assuming narrow width approximation. The bottleneck of a complete NLO computation is the virtual corrections to the interference between the signal and the QCD background, which involve two-loop multiscale integrals that are currently unknown.

In this work we investigate interference effects between the signal $g g \to \Phi \to t \bar{t}$ and background $g g  \to t \bar{t}$, where $\Phi$ represents a spin-0 particle, taking higher order effects into account. Our method can be applied to any UV complete model involving heavy scalar particles. We demonstrate our results in a simplified model with an additional scalar or pseudoscalar particle and the Two-Higgs-Doublet-Model (2HDM). The predictions for the signal and interference can then be compared with the experimental results to obtain constraints on models with new scalars. In particular the impact of higher order QCD effects and of taking into account the interference on the excluded parameter space regions can be explored. Our implementation is available within the {\sc MadGraph5\_aMC@NLO} framework \cite{Alwall:2014hca}. 

This paper is organised as follows. In Section \ref{setup} we discuss the simplified model and 2HDM benchmarks we will employ in our study as well as our computation setup. In Section \ref{features} we explore the main features of top pair production in the presence of additional scalars. Higher order QCD effects including the effect of additional jet radiation and NLO corrections for the process are discussed in Section \ref{higherorder}. In Section \ref{constraints} we examine the impact of our improved predictions on the constraints that can be set on new physics models using top pair resonance searches. The implications of the reported diphoton excess at 750 GeV in the context of top pair production are studied in Section \ref{750}, before we conclude in Section \ref{conclusions}.

\section{Top pair production in the presence of heavy scalars}
\label{setup}
 
\begin{figure}[h!]
\centering
\includegraphics[scale=0.43]{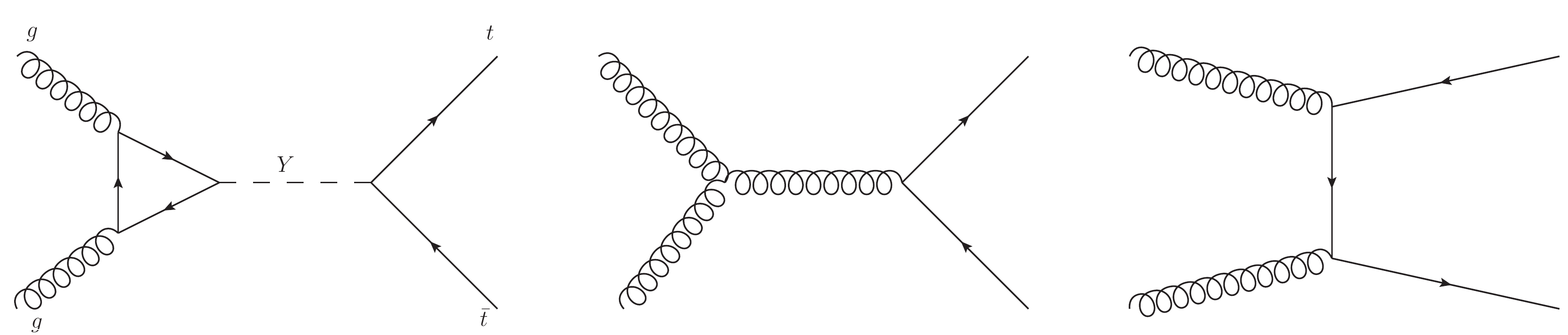}
\caption{Leading order Feynman diagrams for $gg \to t\bar{t}$ production in the presence of spin-0 particles coupling to the top quark. }
\label{diagrams0}
\end{figure}

In the presence of additional scalar particles, the leading order, $\mathcal{O}(g_s^2)$, diagrams for the signal and the  QCD background are shown in figure~\ref{diagrams0}. Any possible CP-even (including the light 125 GeV Higgs), CP-odd or mixed CP scalars are denoted generically by $Y$ in the Feynman diagrams.\footnote{While bottom quarks couple to the SM Higgs and possibly to the heavy scalars, their contribution is very suppressed (for moderate bottom Yukawa couplings) in the region of interest which lies above the top--anti-top threshold.  Therefore we will only consider top quark loops throughout this study.} A couple of observations are in place here. We first note that the signal/background interference is colour suppressed  at leading order in QCD.  The QCD amplitude (for instance the one on the right in figure 1) can interfere with the signal only when the top-quark pair is in a colour singlet (i.e. with probability  $\simeq 1/N_c^2$, $N_c$ being the number of colours).  We also mention that the amplitude for the signal is proportional to the square of the coupling of the scalar to the top, which implies that unless another heavy coloured state runs in the loop, there is no sensitivity to the sign of the Yukawa coupling. In this work we will consider the process of figure \ref{diagrams0} in a simplified model and the 2HDM. The parameters of our BSM models relevant for this process are briefly presented here.  
 
\subsection{Simplified Model}
A simplified model in which one or two additional scalars (one scalar and one pseudoscalar) are present is considered first. Heavier CP-even and CP-odd scalars are denoted by $H^0$ and $A^0$ respectively, coupling to the top in the following way: 
\begin{eqnarray}
\mathcal{L}&=&\bar{t} \frac{y^t}{\sqrt{2}}g^S_{t} t \,H^0+ \bar{t} \frac{y^t}{\sqrt{2}}ig^P_{t} \gamma^5 t \,A^0.
\label{langX0}
\end{eqnarray}
For convenience we normalise the scalar and pseudoscalar interactions with the top quark ($t$) by the SM top Yukawa coupling ($y^t = \frac{m_t}{v}$). We note here that a mixed state $(Y)$ is also allowed in our model, coupling to the top in the following way:
\begin{eqnarray}
\mathcal{L}&=&\bar{t} \frac{y^t}{\sqrt{2}}(g^S_{t}+ig^P_{t} \gamma^5) t \,Y, 
\label{langX0}
\end{eqnarray}
with CP-violation present when both $g^S_{t}$ and $g^P_{t}$ are non-zero.

In this simplified model, the parameters of interest for the production of the heavy scalars and decay into top pairs are: the Yukawa couplings $g_t^S, g_t^P$, the new particle masses $m_H, m_A, m_{Y}$ and their widths $\Gamma_H, \Gamma_A, \Gamma_Y$. A minimum value for the widths can be obtained by computing the partial top decay width and loop-induced (suppressed) decays to gluons and photons through top-quark loops. The parameters of the model can be matched to UV complete models such as the 2HDM.  In particular the total width of the particle can be larger if it couples directly to other SM particles, or new states such as a Dark Matter candidate as explored in \cite{Backovic:2015soa,Mattelaer:2015haa,Arina:2016cqj}. In what follows we will present results for the simplified model using the minimal width and a larger value allowing for other decay channels. 

Finally we note that in our model implementation one can have spin-0 particles directly coupling to the gluons through the following  dimension-5 operators:
\begin{equation}
L_g=\frac{\alpha_s c^S_g}{12\pi v}G_{\mu\nu}G^{\mu \nu} H^0-\frac{\alpha_s c^P_g}{8\pi v}G_{\mu\nu}\tilde{G}^{\mu \nu} A^0. 
\label{gluonoperators}
\end{equation}
This form of interaction is what one obtains in the case of additional heavy coloured states in the loop which can be integrated out, with $c^P_g$ and $c^S_g$ representing the rescaling of the heavy quark Yukawa couplings with respect to that of the top in the SM. In fact, the operators in eq.~\ref{gluonoperators} match the infinite top mass limit used for the SM Higgs when $c^S_g=1$ and $c^P_g=0$. We will employ these operators in Section \ref{750}.

\subsection{The Two--Higgs--Doublet Model}
\label{sec:model} 
To study our process in a UV complete model,  we employ the 2HDM. The 2HDM~\cite{Branco:2011iw} introduces  
a second $SU(2)_L$ doublet $\Phi_2$ and gives rise to five physical
Higgs bosons: one light (heavy) neutral, CP-even state $\hzero$ ($\Hzero$); 
one neutral, CP-odd state $\Azero$; and two charged Higgs bosons $H^{\pm}$.
In this work we identify $\hzero$ with the Higgs particle
observed at the LHC and fix its mass to $m_{\hzero} = 125$ GeV.
The free input parameters determining all properties of a 2HDM scenario are:
\begin{alignat}{9}
 \tan\beta\;, \sin\alpha\;, m_{\hzero}\;, m_{\Hzero}\;, m_{\Azero}\;, m_{H^{\pm}}\;, m^2_{12}
 \label{eq:freeinputs}.
\end{alignat}

The convention
$0 \leq \beta-\alpha < \pi$ (with $0 < \beta < \pi/2$) ensures 
that the Higgs coupling to the weak gauge bosons $g^{\text{2HDM}}_{hVV} = \sin(\beta-\alpha)\,g^{\text{SM}}_{hVV}$ has the same sign as in the SM. 

Two types of 2HDM scenarios can arise i) type--I, in which 
all fermions couple to just one of the Higgs
  doublets; and ii) type--II, where up--type (down--type) fermions couple
 only to $\Phi_2$ ($\Phi_1$). 
The resulting Yukawa couplings deviate from the SM ones by:
\begin{alignat}{9}
g_{hxx} &\equiv g^h_x  = 
\left( 1 + \Delta^h_x \right) \; g_x^\text{SM} \; .
\label{eq:delta}
\end{alignat}  
 Analytic expressions for these coupling shifts for the top are provided 
in table~\ref{tab:yukawas} for both type--I and type--II 2HDMs. 
 
 \begin{table}[t!]
 \centering
\begin{tabular}{l|l}
  \multicolumn{2}{c}{Type--I and II}  \\ \hline
 $1+\Delta^{\hzero}_t$ & $\cfrac{\cos\alpha}{\sin\beta} $  \\
 $1+\Delta^{\Hzero}_t$ & $\cfrac{\sin\alpha}{\sin\beta}  $  \\
  $1+\Delta^{\Azero}_t$ & $\cot \beta$  \\
\end{tabular}
\caption{Top quark Yukawa couplings to the light (heavy)
 CP-even and CP-odd Higgs bosons. These are identical for type--I and type--II 2HDM. \label{tab:yukawas}}
\end{table}

 Electroweak precision tests, the LHC Higgs results and searches for heavy scalar particles, along with unitarity, perturbativity
and vacuum stability constrain the parameter space of the model. In the selection of 2HDM benchmarks, these constraints are taken into account. 

Our set of representative 2HDM benchmark scenarios are introduced in table~\ref{tab:benchmarks}. These are all type--II and have been constructed in  agreement with all up--to--date parameter
space constraints, which we have included through  the public tools { \sc 2HDMC}~\cite{Eriksson:2009ws},
{ \sc HiggsBounds}~\cite{\higgsbounds}, 
{\sc SuperIso}~\cite{\superiso} and {\sc{HiggsSignals}} \cite{Bechtle:2013xfa,Stal:2013hwa}. 
We note here that as discussed in \cite{Mahmoudi:2009zx}, values of $\tan \beta < 1$ (resulting in enhanced top Yukawa couplings) are ruled out for a type--I 2HDM by $BR(B \to X_s \gamma)$ and $\Delta M_{B_d}$ experimental constraints. On the other hand, for a type--II 2HDM  $\tan \beta < 1$ is allowed as long as $m_{H^\pm} > 600$ GeV. 
In table~\ref{tab:couplings} we quote the numerical values
for the top Yukawa couplings, scalar widths and top branching ratios for the benchmarks 
defined in table~\ref{tab:benchmarks}. 
All couplings
are normalized to their SM counterparts, as denoted
by $\hat{g}_{hxx} \equiv g^{\text{2HDM}}_{hxx}/g^{\text{SM}}_{Hxx}$,
where $H$ stands for the SM Higgs boson. 

\begin{table}[tb!]
\begin{center}
 \begin{tabular}{|l||rr|rrrr|} \hline 
 & $\tan\beta$ & $\alpha/\pi$ & $m_{\Hzero} $ &  $m_{\Azero} $  & $m_{H^{\pm}} $   & $m^2_{12} $  \\ \hline
B1 & 1.75 & -0.1872 & 300 & 441 & 442 & 38300 \\
B2 & 0.9 & -0.267 & 500 & 550 & 620 & 10000 \\
B3 & 0.7 & -0.306 & 380 & 590 & 610 & 10000 \\
B4 & 0.6 & -0.328 & 500 & 710 & 720 & 10000 \\ \hline
 \end{tabular}
 \end{center}
 \caption{Parameter choices for the different 2HDM benchmarks used in our study. All
 masses are given in GeV. The lightest Higgs mass is fixed in all cases to $m_{\hzero} = 125$ GeV.  \label{tab:benchmarks}}
\end{table}

\begin{table} [tb!]
\begin{center}
    \begin{tabular}{ | l || c | c | c | c | c | c | c |}
    \hline 
      & $\hat{g}_{\hzero tt}$ & $\hat{g}_{\Hzero tt}$ & $\hat{g}_{\Azero tt}$ & $\Gamma_{\Hzero}$ (GeV) & $BR(H^0\to t\bar{t})$ & $\Gamma_{\Azero}$ (GeV)  & $BR(A^0\to t\bar{t})$   \\ \hline 
B1 & 0.96 & -0.64 & 0.57 & 0.138 & 0.0 & 7.20 & 0.723 \\
B2 & 1.00 & -1.11 & 1.11 & 13.75 & 0.9997 & 29.97 & 0.9997 \\
B3 & 1.00 & -1.43 & 1.43 & 3.39 & 0.9989 & 64.57 & 0.849 \\
B4 & 1.00 & -1.67 & 1.67 & 30.93 & 0.9998 & 105.23 & 0.896 \\ \hline  
   \end{tabular}
\end{center}
\caption{Normalised top quark Yukawa couplings, heavy scalar widths and top branching ratios for the
different 2HDM benchmarks defined in table~2. All couplings are normalised to 
their SM counterparts. \label{tab:param}}
\label{tab:couplings}
\end{table}

\medskip{}
The properties of the different 2HDM scenarios can be summarised as follows:
\begin{itemize}
 \item{\textbf{B1:}} The $\tan \beta >1$ is responsible for smaller top Yukawa couplings for the heavy scalars. The CP-even scalar has a rather narrow width (main decay channel is $h^0h^0$) and lies below the resonant top--anti-top threshold while for the pseudoscalar the branching ratio to tops rises to more than 70$\%$.
 \item{\textbf{B2: }} Both new resonances feature slightly enhanced top Yukawa couplings. In this scenario the new particles are all around 500 GeV and always decay into $t\bar{t}$. The widths of the heavy scalars remain below 10\% of the mass.  
 \item{\textbf{B3:}} The top Yukawa couplings are enhanced due to the smaller value of tan $\beta$. Compared to B2 the width of the scalar is suppressed due to its lower mass while the width of the pseudoscalar reaches $\sim$10\% of its mass. The CP-even state decays almost exclusively to top quarks, while $\Azero$ can also decay into a ZH pair. 
  \item{\textbf{B4:}} Both resonances are rather broad with a larger mass hierarchy and enhanced couplings to the top quark. Both $\Hzero$ and $\Azero$ decay predominantly to top quarks. 
\end{itemize}

\subsection{Technical setup}

In this work, we employ the {\sc MadGraph5\_aMC@NLO} framework \cite{Alwall:2014hca}. The one-loop amplitudes  have been obtained with {\sc MadLoop} \cite{Hirschi:2011pa} by computing one--loop matrix elements using the {\sc OPP} integrand--reduction method~\cite{Ossola:2006us} (as implemented in {\sc CutTools}~\cite{Ossola:2007ax}). Signal events are generated at LO with the latest {\sc MadGraph5\_aMC@NLO} \cite{Hirschi:2015iia}, which allows event generation for loop-induced processes.  For the interference and the signal at NLO, a reweighting procedure has been followed. Reweighting has been employed already for a series of processes within the {\sc MadGraph5\_aMC@NLO} framework \cite{Frederix:2014hta,Hespel:2014sla,Maltoni:2014eza,Frederix:2016cnl} both at LO and NLO accuracy and has recently been automated and made public as part of the official code release \cite{Olivier}. This procedure involves generating events through the implementation of a tree-level effective field theory, using a UFO model \cite{Degrande:2011ua,deAquino:2011ub}. After event generation, event weights obtained from the tree-level EFT amplitudes are modified by the ratio of the full one-loop amplitude over the EFT ones, {\it i.e.}, $r=|\mathcal{M}_{\textrm{Loop}}^2|/|\mathcal{M}_{\textrm{EFT}}^2|$, where $|\mathcal{M}_{\textrm{Loop}}^2|$ represents the numerical amplitude as obtained from {\sc MadLoop}.  For the NLO computation weights corresponding to Born, virtual corrections and real emission configurations are reweighted using the corresponding matrix element.  More details of the reweighting specific to the NLO computation of the signal will be presented in Section \ref{NLO}.

\section{Features of additional scalar contribution to top pair production}
\label{features}

In this section we explore the main features of the top pair production process in the presence of new scalars. Conclusions can be drawn already at the amplitude squared level by varying the various model parameters. We investigate the interference patterns between the signal and the QCD background in the presence of 
\begin{itemize}
\item  one state, CP-even or CP-odd, 
\item  one CP-mixed state 
\item two states, one CP-even and one CP-odd. 
\end{itemize}
We note here that while both scalar and pseudoscalar amplitudes interfere with the QCD background, there is no interference between them if the two spin-0 states are one pure CP-even and one pure CP-odd. We also mention that bottom quarks in principle enter in this process as they have non-zero Yukawa couplings.  However, as the effects we consider here concern the region above the top--anti-top threshold, any contribution of bottom quarks is expected to be very small for moderate values of the bottom Yukawa. In the 2HDM scenarios we consider a small value of tan$\beta$, so that all bottom Yukawa couplings are small. In fact in the 2HDM, when the top Yukawa coupling is increased,  which is what our benchmarks aim to do, then the bottom Yukawa coupling is automatically reduced. In conclusion, we can safely ignore bottom-quark loops in what follows. 

In the case of a CP-even or mixed state the signal interferes also with the SM-like Higgs (125 GeV) contribution. We find that this interference is suppressed compared to the interference with the QCD background yet we do include it in our results. We also compute the light Higgs contribution to the SM background, both the pure Higgs contribution and its interference with the QCD background. Both are extremely  suppressed compared to the QCD background. 

\begin{figure}[h!]
\centering
\includegraphics[scale=0.54]{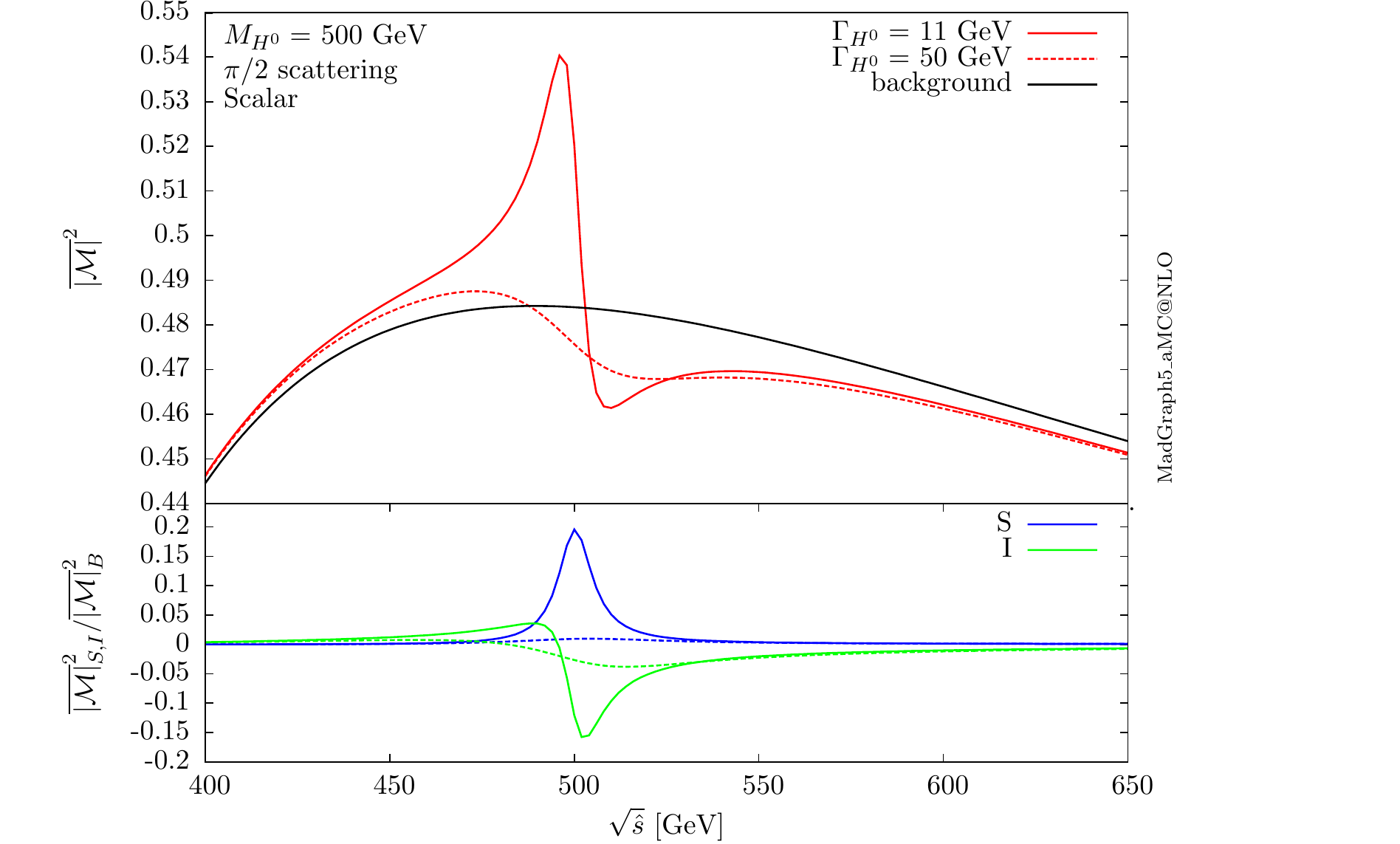}
\includegraphics[scale=0.54]{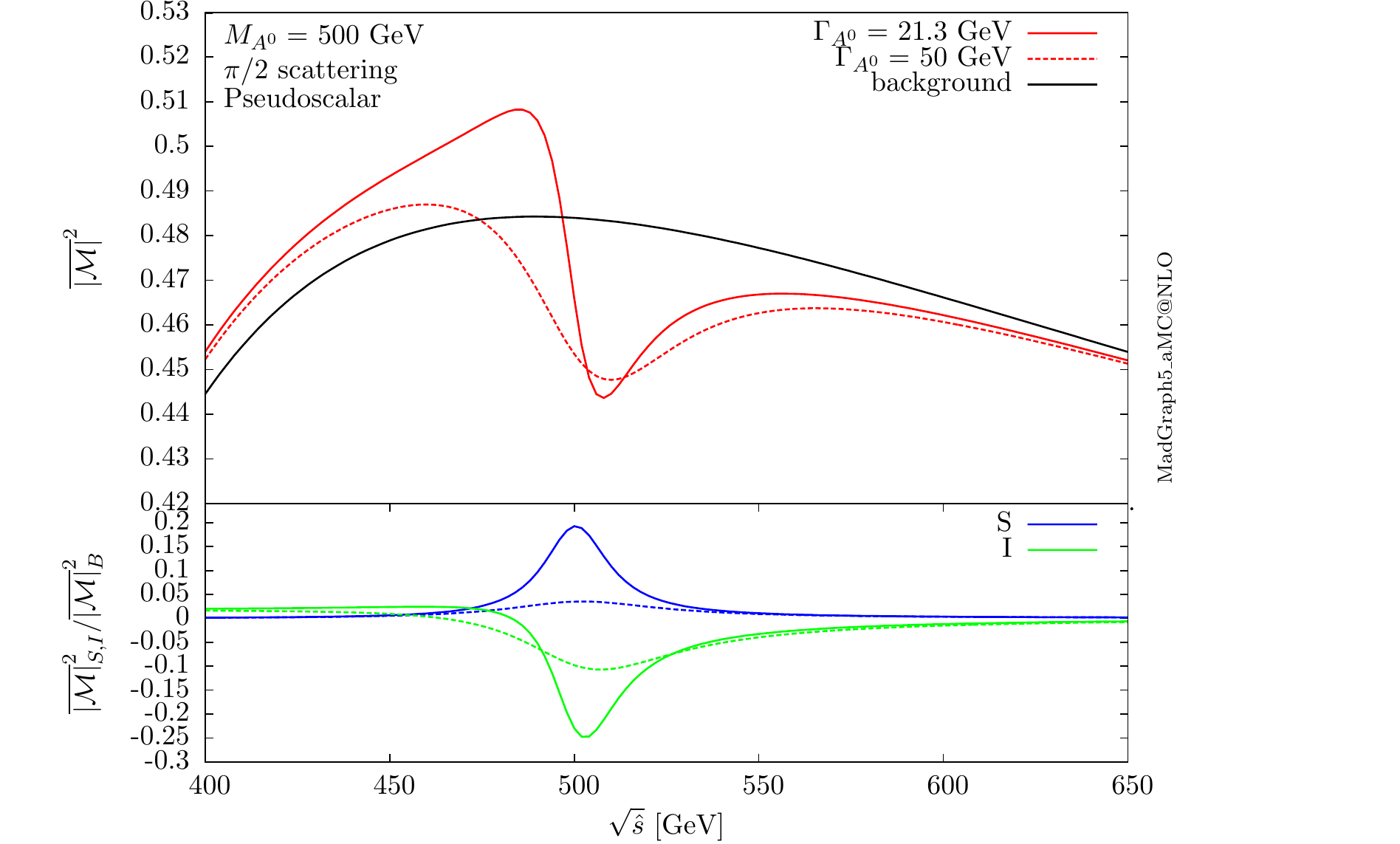}
\includegraphics[scale=0.54]{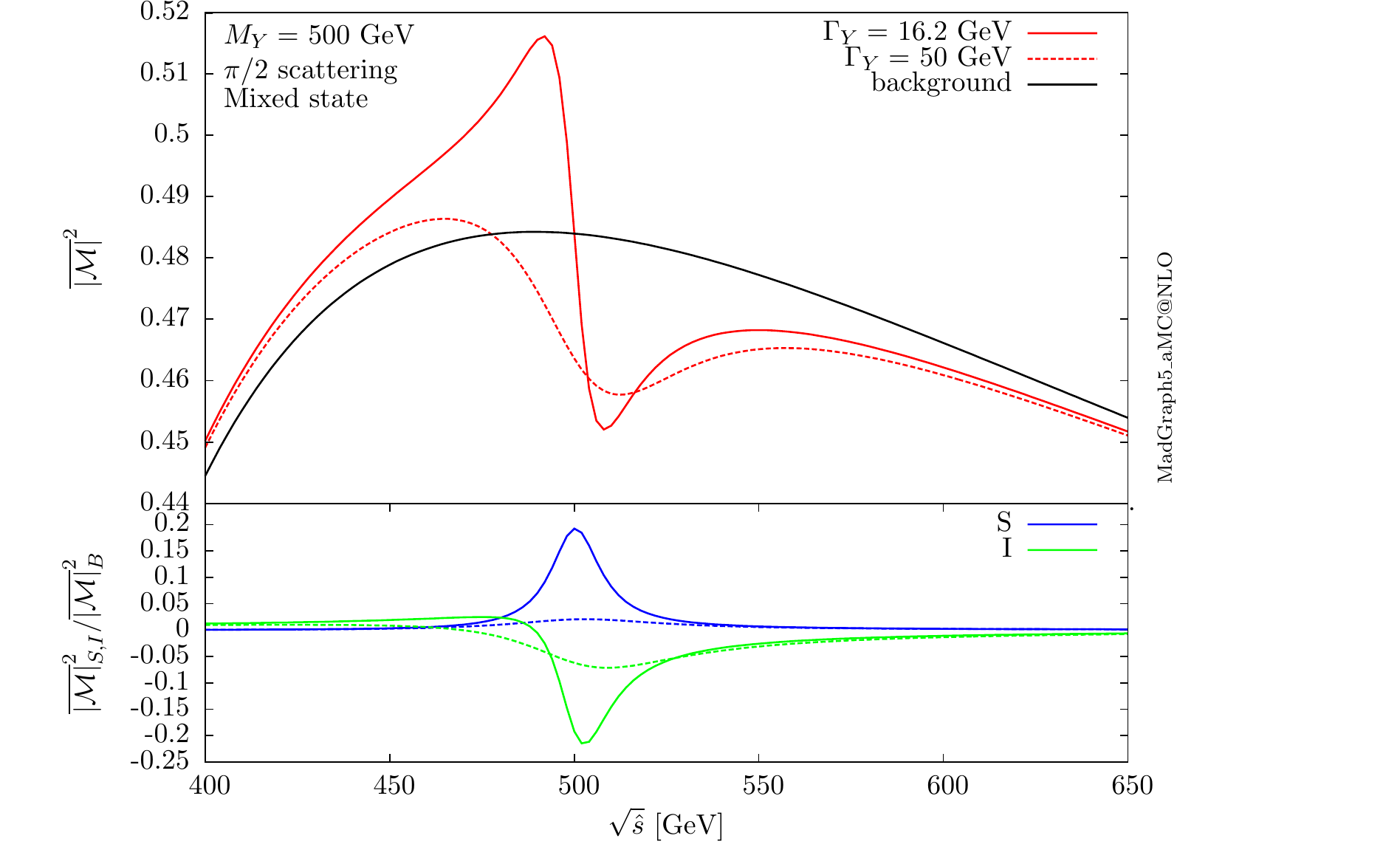}
\caption{Amplitude squared for $gg (\to \Phi)\to t \bar{t}$ in the presence of a heavy scalar of 500 GeV as a function of the centre-of-mass energy for different widths. The centre-of-mass frame scattering angle is set to $90^\textrm{o}$. Results are shown for a scalar state (top), a pseudoscalar one (centre) and a mixed one (bottom). The lower inset shows the ratio of the signal and interference over the QCD background. }
\label{total_HH}
\end{figure}

To demonstrate our results we select the invariant mass distribution of the top pair, an observable which can very clearly reveal the presence of a resonance.\footnote{We note here that top decays can also be generated in our simulation framework. While observables involving top decay products are known to  provide useful information on the nature of a top resonance \cite{Frederix:2007gi}, in this work for brevity we will only consider stable top quarks.} As an example we show in figure~\ref{total_HH} the amplitude squared for the signal, background and interference separately for a scalar, a pseudoscalar  and mixed (equal scalar and pseudoscalar couplings) spin-0 state for various widths (90$^{\textrm{o}}$ centre-of-mass frame scattering angle). The Yukawa couplings ($g^{S,P}_t$) are all set to 1. The values of the widths chosen for the plots are i) the minimal width computed at LO assuming the scalar particle only couples to the top and ii) a larger width to allow for decays to other SM particles (e.g. vector bosons) or new states (such as a Dark Matter particle). The plots show that the interference is important even for narrow resonances with widths as small as $\sim$2\% of the mass, which is the case for the scalar 500 GeV resonance. The interference can be as large as the signal in size and leads to the characteristic peak-dip structure. The different width choices highlight the impact of the width of the additional particle on the relative importance of the interference. When the width of the heavy scalar becomes large ($\sim 10$\% of the mass) the peak-dip structure becomes less pronounced and basically leads to a dip dominated by the interference. Note also that the pseudoscalar resonance peak reaches larger values than the scalar case for same mass and width, which is related to the structure of the top loop amplitudes for $gg\to H^0/A^0$.  

\begin{figure}[h!]
\includegraphics[scale=0.50,trim=2cm 0 0 0]{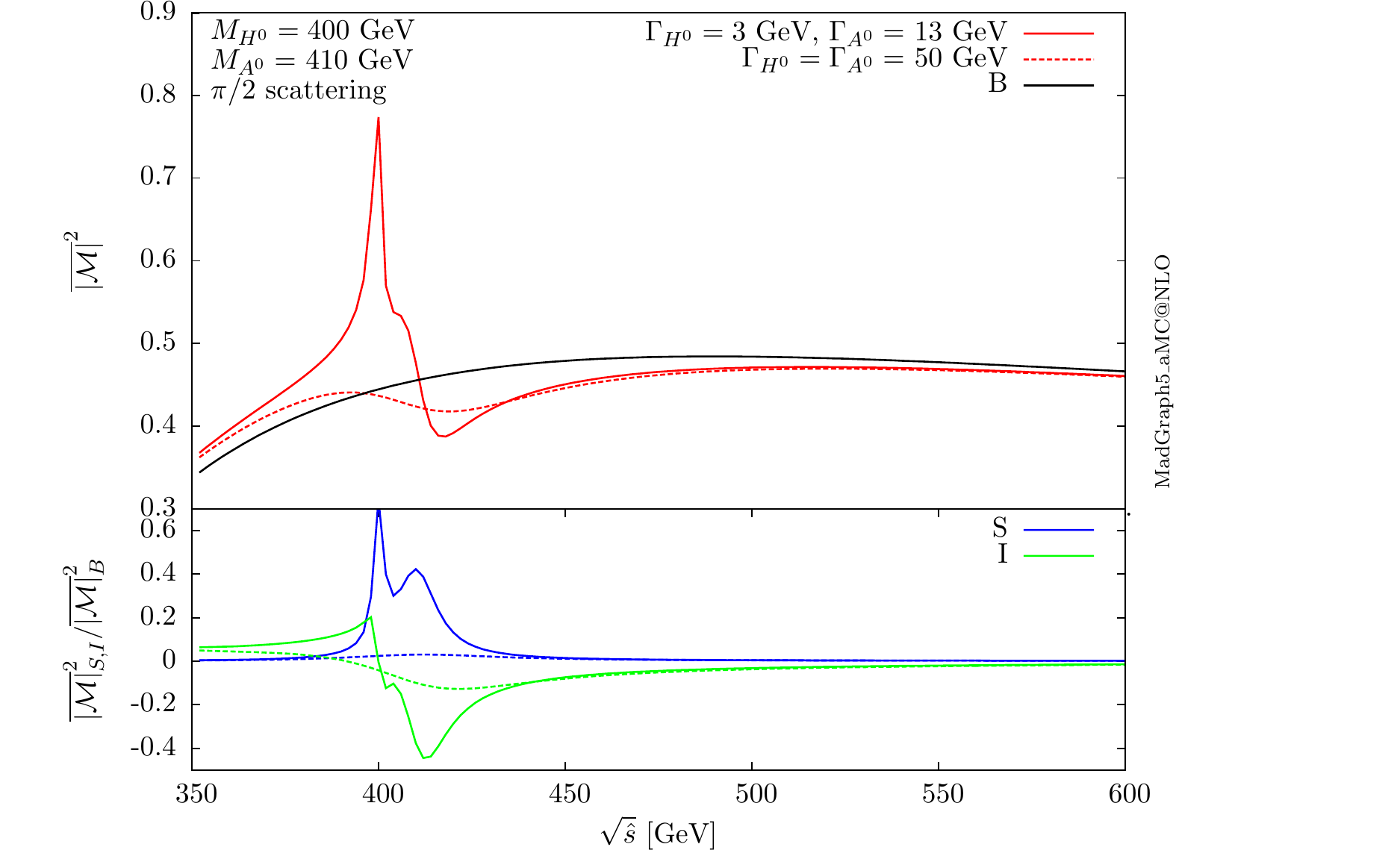}
\includegraphics[scale=0.50, trim=3cm 0 0 0]{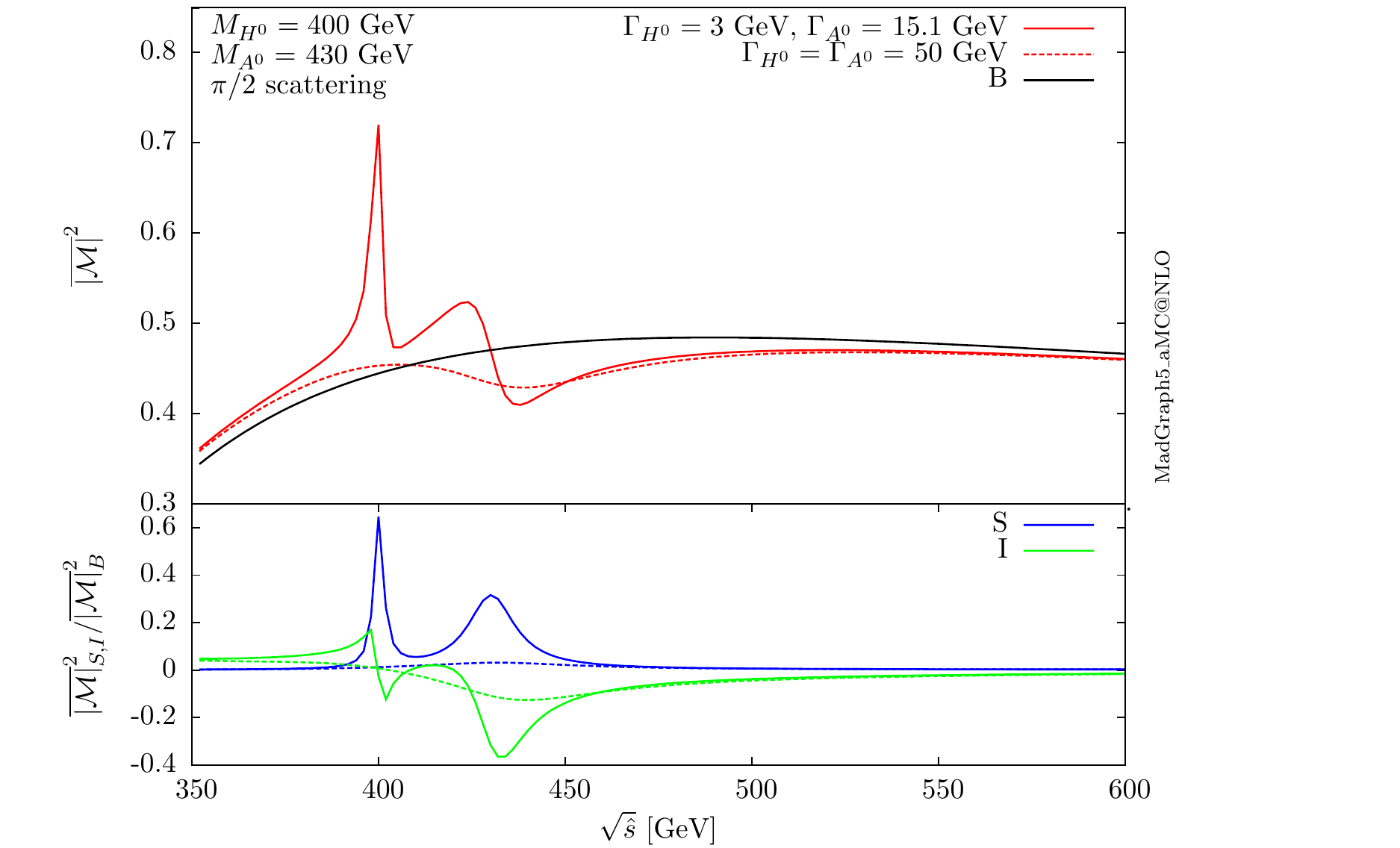}
\includegraphics[scale=0.50,trim=2cm 0 0 0]{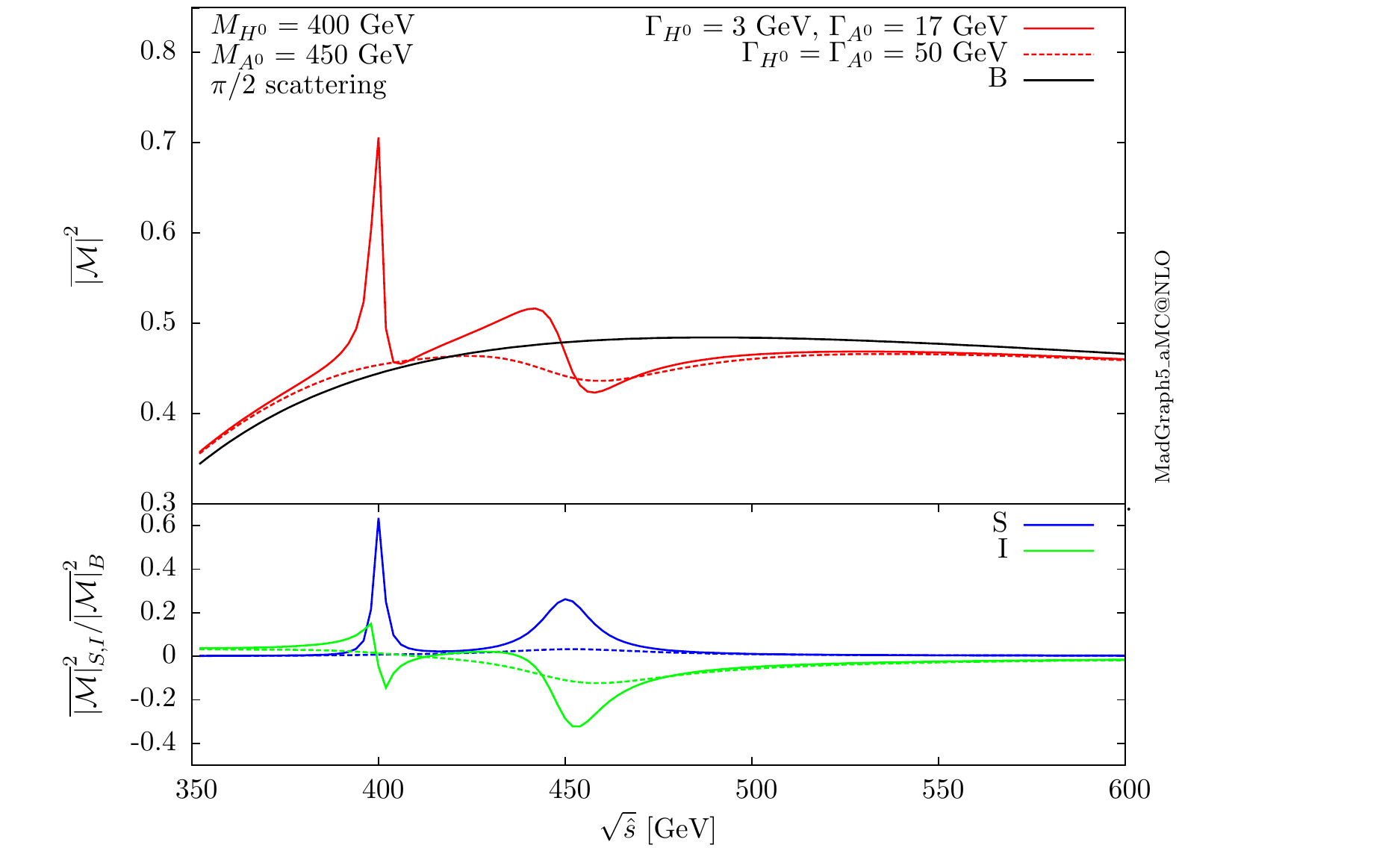}
\includegraphics[scale=0.50, trim=3cm 0 0 0]{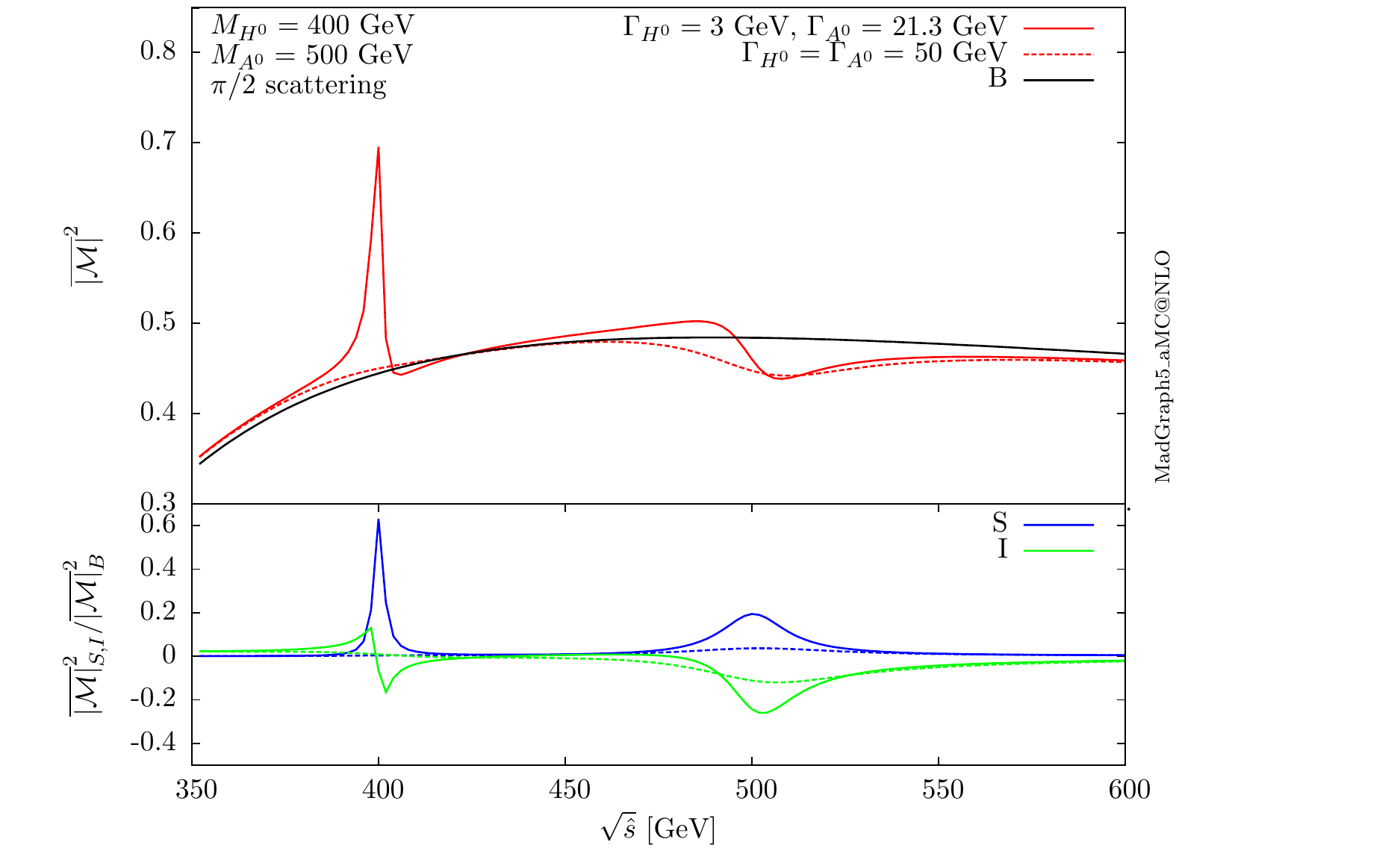}
\caption{Amplitude squared for $gg(\to \Phi)\to t \bar{t}$ in the presence of a scalar and a pseudoscalar state as a function of the centre-of-mass energy for different mass splittings and widths. The centre-of-mass frame scattering angle is set to $90^\textrm{o}$.  Top left: small mass splitting ($\Delta m=10$ GeV),  top right: moderate mass splitting ($\Delta m=30$ GeV), bottom left: large mass splitting ($\Delta m=50$ GeV) and bottom right: larger mass splitting ($\Delta m=100$ GeV). All couplings are equal to the SM top Yukawa. The lower inset shows the ratio of the signal and interference over the QCD background.}
\label{total_all}
\end{figure}

In a scenario where both a heavy scalar, $H^0$, and a pseudoscalar, $A^0$, are present, more interesting features arise in the invariant mass distribution of the top pair as discussed also in \cite{Jung:2015gta}. We consider this scenario in figure~\ref{total_all}, where the amplitude squared is studied in the presence of one scalar and one pseudoscalar particle. The patterns observed in the invariant mass distribution are determined by the mass splitting and widths of the two particles. In the narrow width case, for $\Delta m=10$ GeV it is not possible to disentangle the two peaks. However, a larger mass splitting leads to two distinctive peak-dip structures. For the 50 GeV widths the effects are very mild compared to the background and dominated by the interference. In practice, the experimental top invariant mass resolution will determine the mass gap required for the two states to be distinguished even in the narrow width case.

\begin{figure}[h!]
\centering
\includegraphics[scale=0.7]{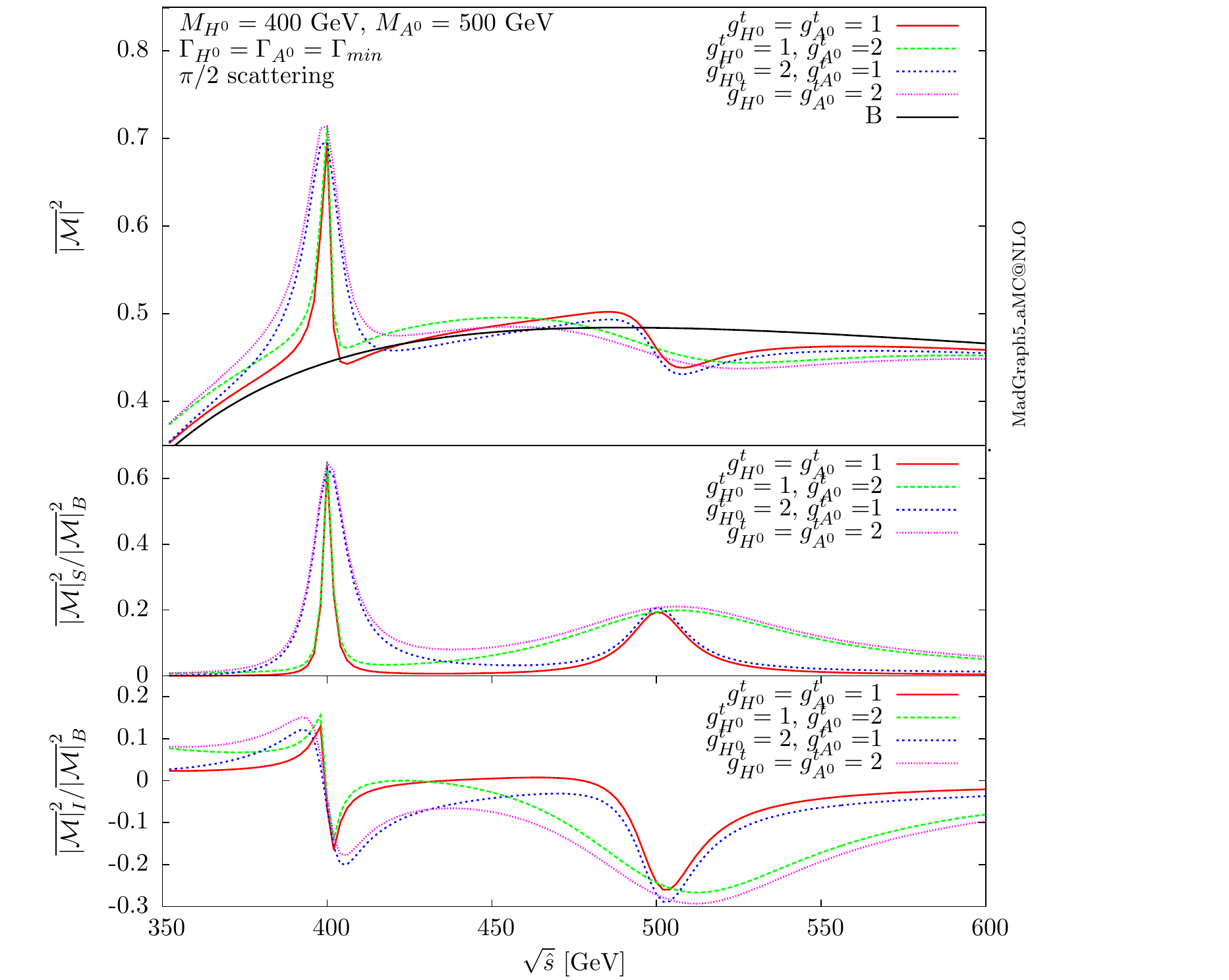}
\caption{Amplitude squared for $gg(\to \Phi) \to t \bar{t}$ as a function of the centre-of-mass energy for the $\Hzero$ and $\Azero$ resonances for different values of the Yukawa couplings. The widths are set to the minimum ones: for $g^t=1, \Gamma_{H^0} = 3$ GeV, $\Gamma_{A^0} = 21.3$ GeV and for $g^t=2, \Gamma_{H^0} = 12$ GeV, $\Gamma_{A^0} = 85.1$ GeV. }
\label{total_Yuk}
\end{figure}

To conclude our amplitude analysis of interference effects in top pair production, we modify the Yukawa couplings $g_t^S$, $g_t^P$ of the new particles. The results are shown in figure~\ref{total_Yuk}. Note that flipping the sign of the Yukawas has no effect on the interference pattern for this process as $\mathcal{M}_{signal} \propto y_t^2$.  In this case, the width is computed at LO assuming only top decays for all values of the Yukawa couplings. The plot demonstrates the range of possible shapes one can expect in the case of two resonances with different signal strengths. The values of the Yukawa couplings change not only the normalisation but also the shape as the interference and signal have different functional dependences on the Yukawa coupling. We also note that the signal and interference cannot simply be rescaled when one changes the Yukawa couplings, as for consistency the width of the heavy state should be appropriately recomputed. In particular we notice that in this model the width rapidly increases with the Yukuwa couplings, quickly reaching values beyond the narrow width approximation.

The same qualitative conclusions can be drawn by studying the results at the proton-proton cross-section level. Leading order results are presented for both the simplified model and the 2HDM scenarios presented above. All results are obtained for the LHC at $\sqrt{s}=13$ TeV with MMHT2014LO PDFs \cite{Harland-Lang:2014zoa}. The renormalisation and factorisation scales are set to $\mu_F=\mu_R=\mu_0=m_{t\bar{t}}/2$. 
The cross sections obtained for the signal and interference at LO are shown in table \ref{table:xsec1} for a scalar and pseudoscalar resonance of 500 GeV for two width choices. For comparison the LO QCD background (including the quark--anti-quark contribution) is $\sigma_{QCD} = 498.1^{+31.4\%}_{-22.4\%}$ pb and the interference between SM Higgs and QCD background is $\sigma_{h-QCD} = -0.90^{+32.4\%}_{-23.2\%}$ pb. In the following when we refer to background we will use $\sigma_{background}^{LO} = \sigma^{LO}_{QCD} + \sigma^{LO}_{h-QCD} = 497.2^{+31.4\%}_{-22.4\%}$ pb. Note that $\sigma_h = \sigma^{pp \to h \to t\bar{t}} = 22.15^{+33.3\%}_{-23.5\%}$ fb is part of the SM background but since its contribution is very small it is here discarded. 

\begin{figure}[h!]
\includegraphics[scale=0.5,trim=2cm 0 0 0]{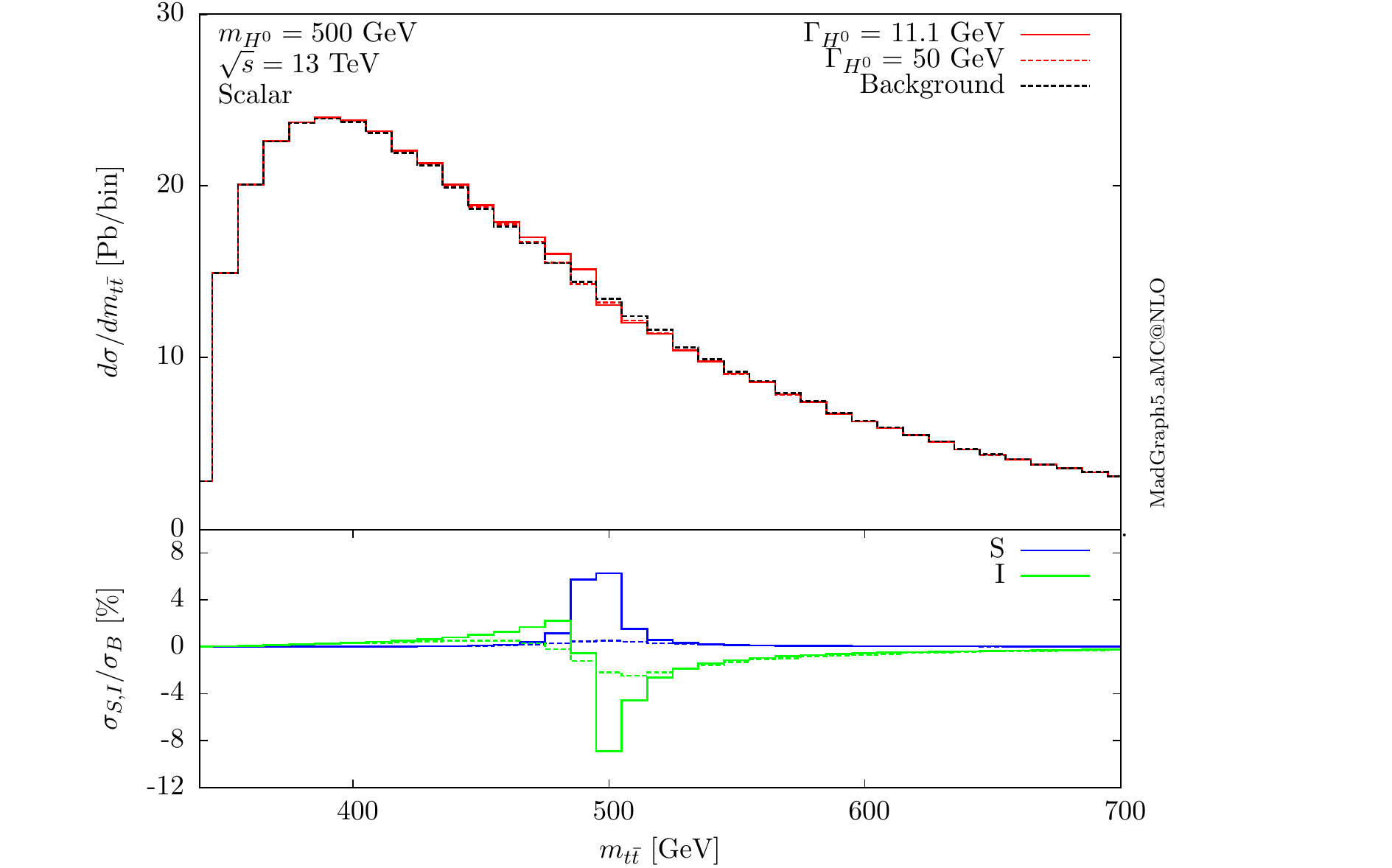}
\includegraphics[scale=0.5,trim=3cm 0 0 0]{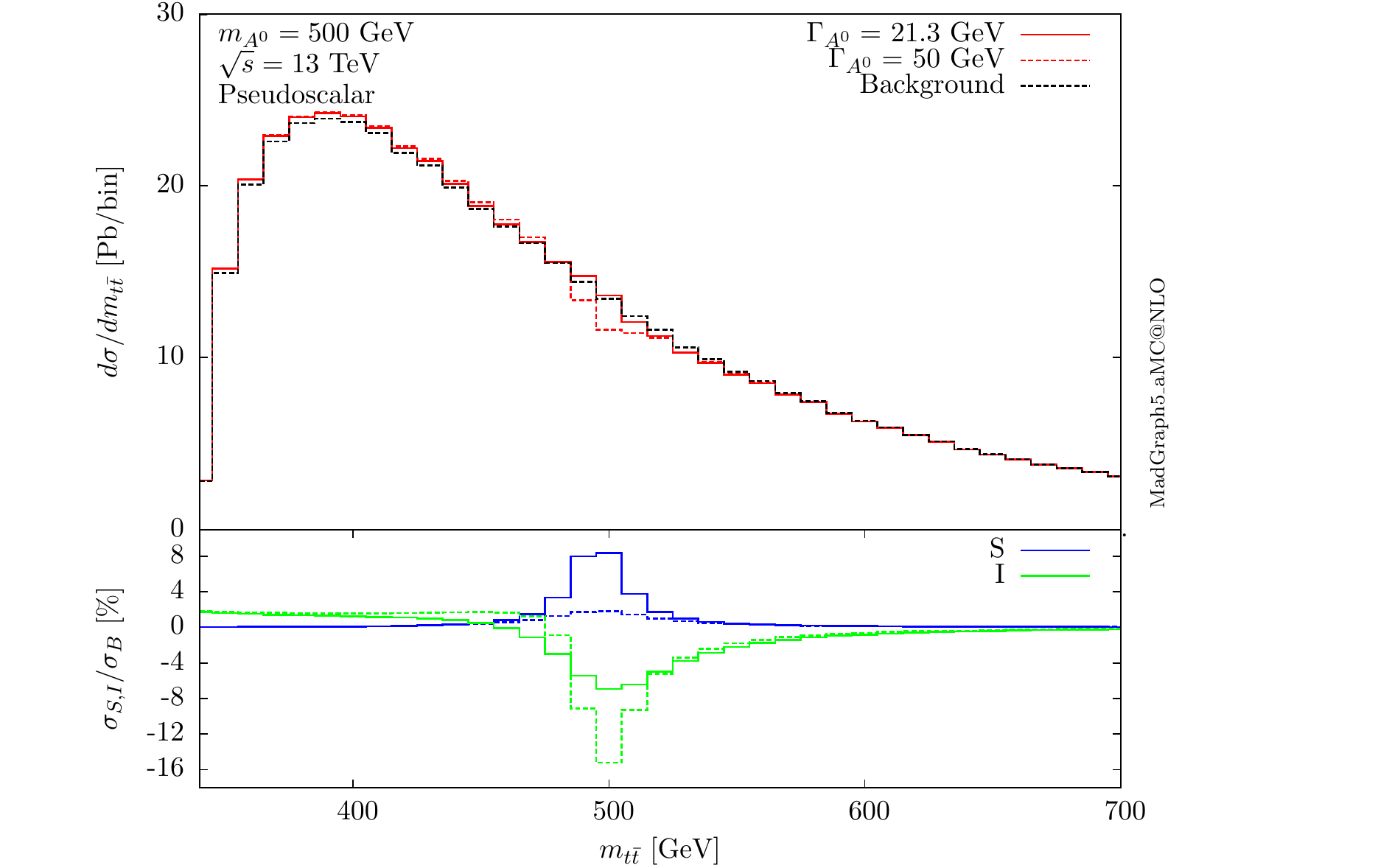}
\caption{Top-pair invariant mass distribution for a single heavy resonance of $m=500$ GeV with different widths at the LHC at 13 TeV. Left: scalar. Right: pseudoscalar. The contributions of the signal and interference are shown separately as a percentage of the QCD background in the lower panels. }
\label{LO_HH_simp}
\end{figure}

Figure~\ref{LO_HH_simp} shows the top invariant mass distribution for a scalar and pseudoscalar state separately and confirms our observations at the amplitude squared level. The effect of the scalar particle remains at the few percent level compared to the background and hardly visible especially due to a cancellation between the signal and interference contributions, as reported in table \ref{table:xsec1}.
From table \ref{table:xsec1} we note that the signal changes with the width following a $\propto1/\Gamma$ behaviour at the total cross-section level, as expected in the narrow-width approximation. Effectively, increasing the total width without changing the partial top width decreases the branching ratio ($\Gamma_{t\bar{t}}/\Gamma$). The impact of changing the width on the interference is not straightforward to predict at the total cross-section level. The interference is decomposed into a part coming from the imaginary part of the one-loop amplitude which is always destructive and one coming from the real part which changes sign at the mass of the resonance, as also discussed in \cite{Djouadi:2016ack}. The total interference can be negative or positive depending on the relative size of these two components. In general though the relative importance of the interference compared to the signal is larger for larger widths as shown in table \ref{table:xsec1}.

\begin{table}[h!]
\begin{center}
    \begin{tabular}{|c|c|c|c|c|}
        \hline 
& \multicolumn{2}{c|}{Scalar} &  \multicolumn{2}{c|}{Pseudoscalar}  \\ \hline
Width &$\Gamma_{min}=11.1$ GeV&$\Gamma = 50$ GeV & $\Gamma_{min}=21.3$ GeV & $\Gamma = 50$ GeV \\ \hline
%Total & 496.74 & 494.84  \\ \hline
Signal & 2.38 & 0.47& 4.54 & 1.81 \\ 
\hline
Interference & -1.27 & -1.25 & -2.19 & -2.50 \\ 
\hline 
\end{tabular}
 \caption{Cross sections (in pb) for the LHC at 13 TeV for the signal and interference with the background for a new heavy scalar or pseudoscalar particle  of $m_{H^0,A^0}=500$ GeV for different width values. Yukawas are equal to the SM values.
\label{table:xsec1}}
\end{center} 

\end{table}

\begin{figure}[h!]
\includegraphics[scale=0.5,trim=2cm 0 0 0]{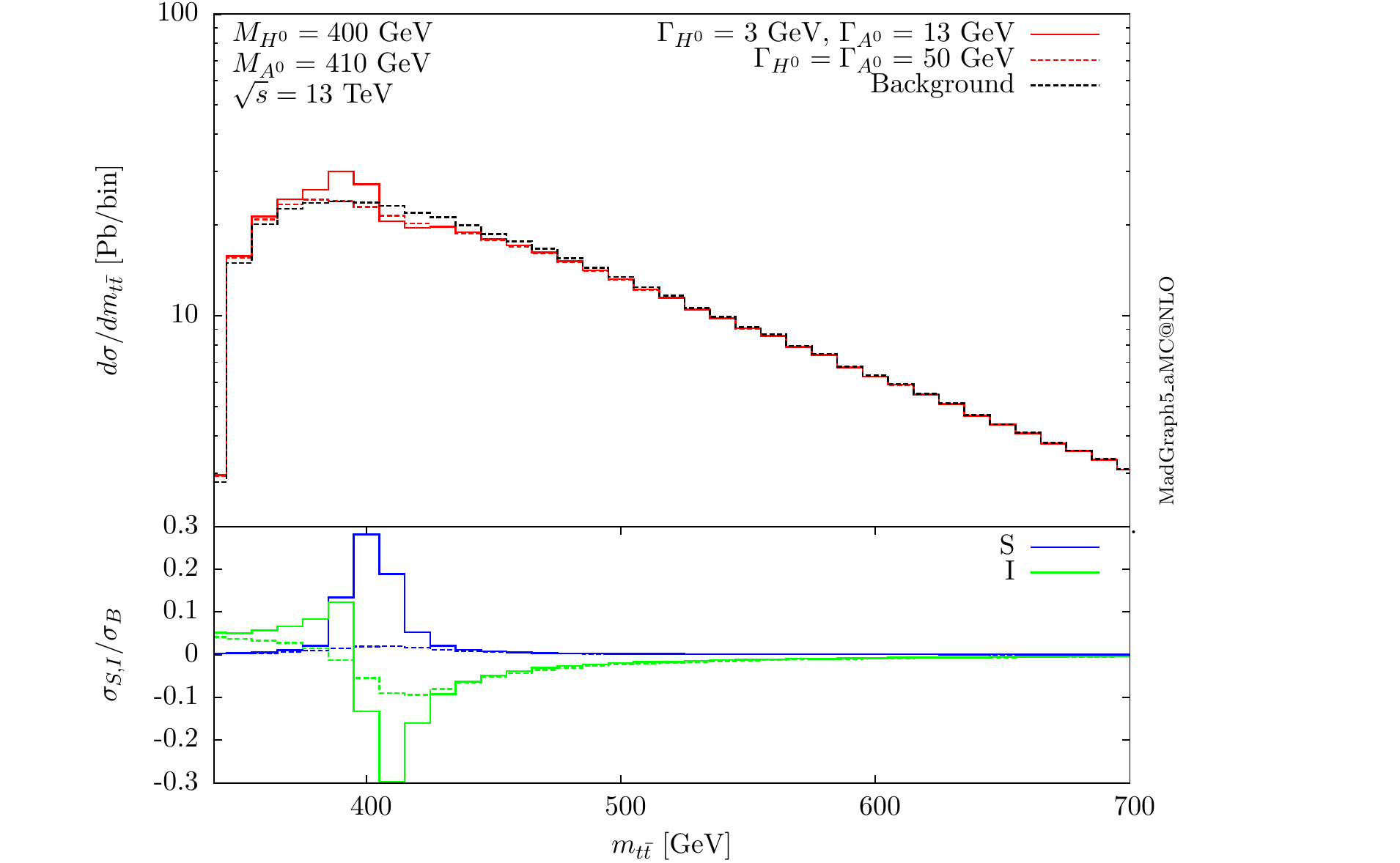}
\includegraphics[scale=0.5,trim=3cm 0 0 0]{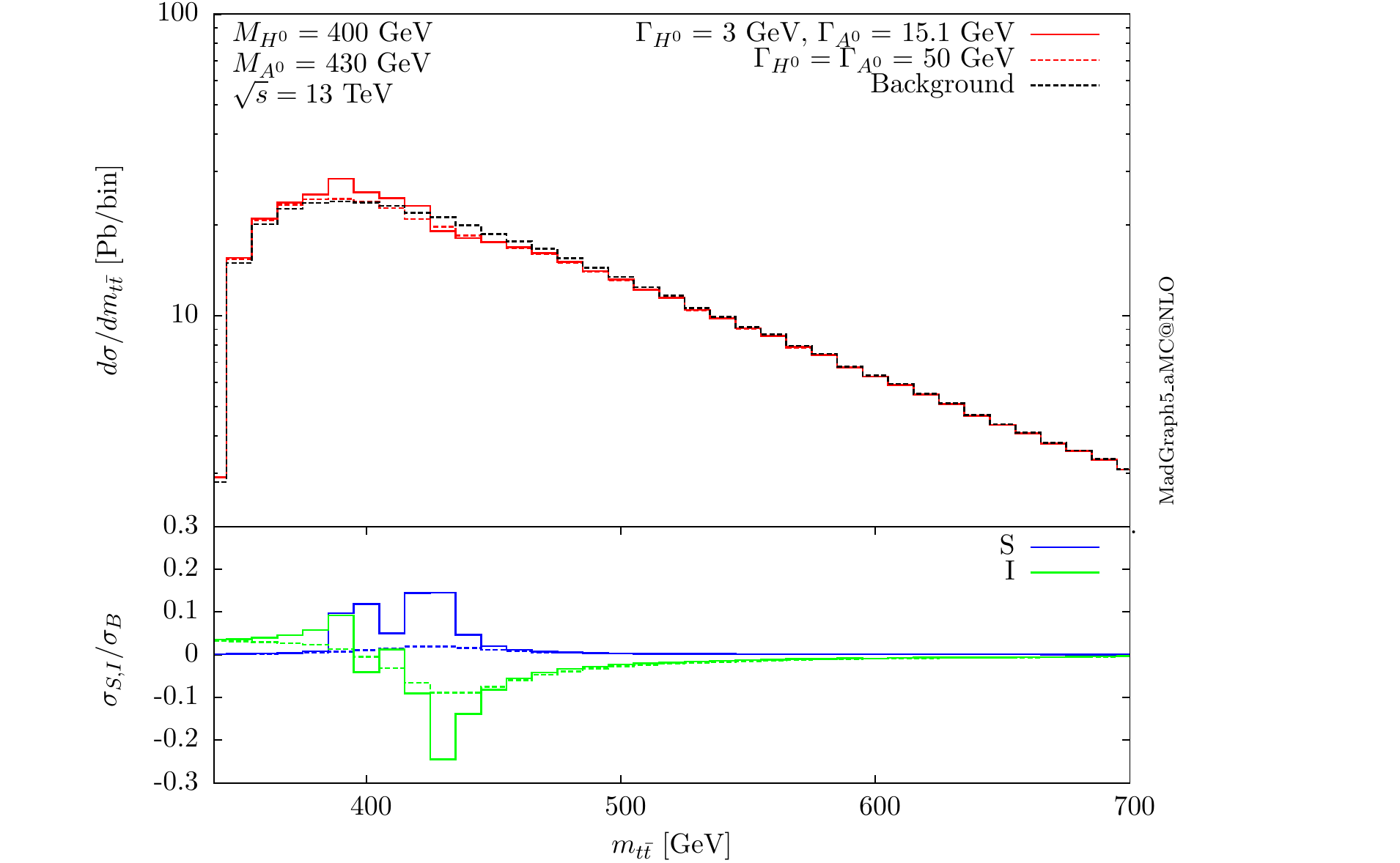}
\includegraphics[scale=0.5,trim=2cm 0 0 0]{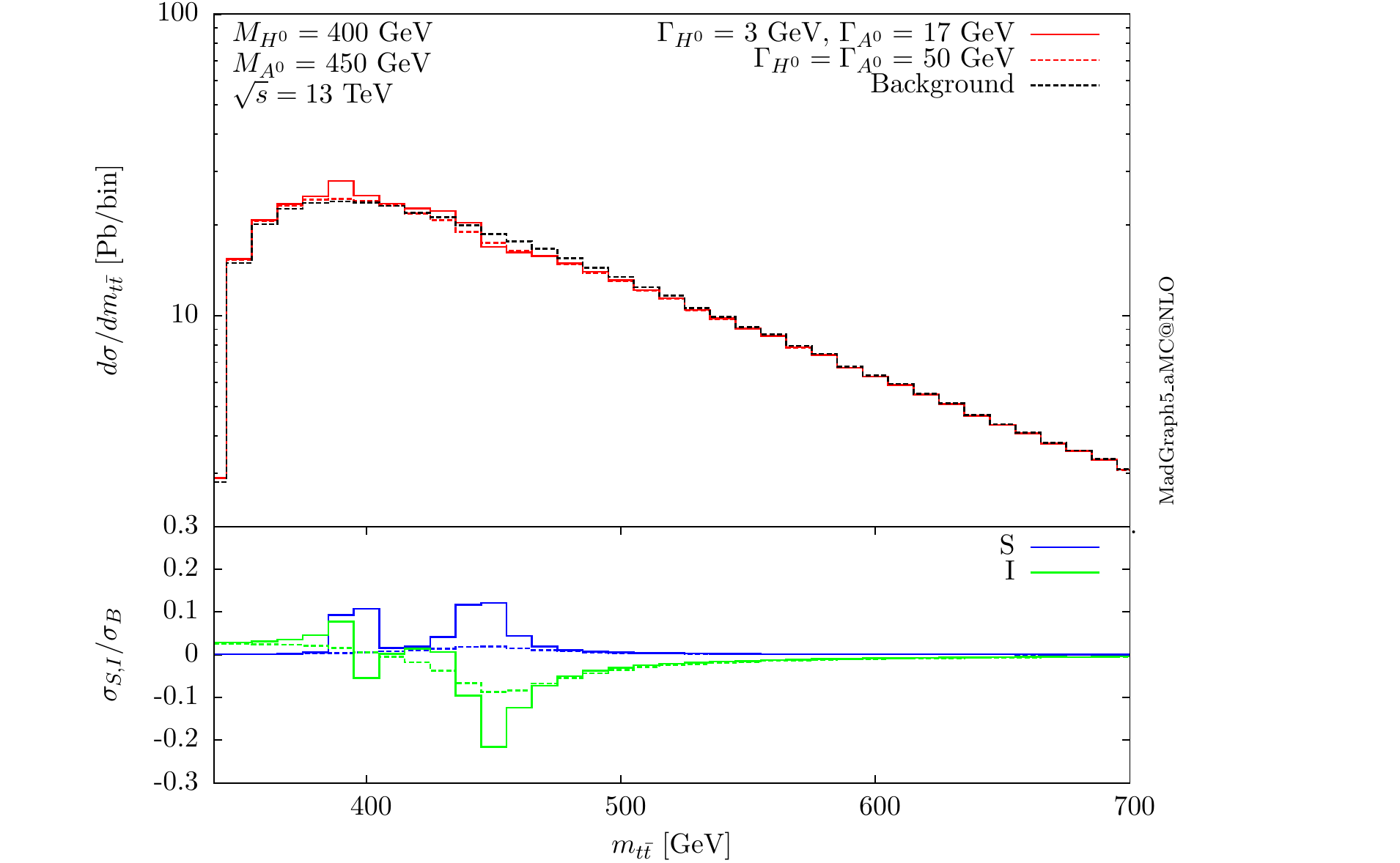}
\includegraphics[scale=0.5,trim=3cm 0 0 0]{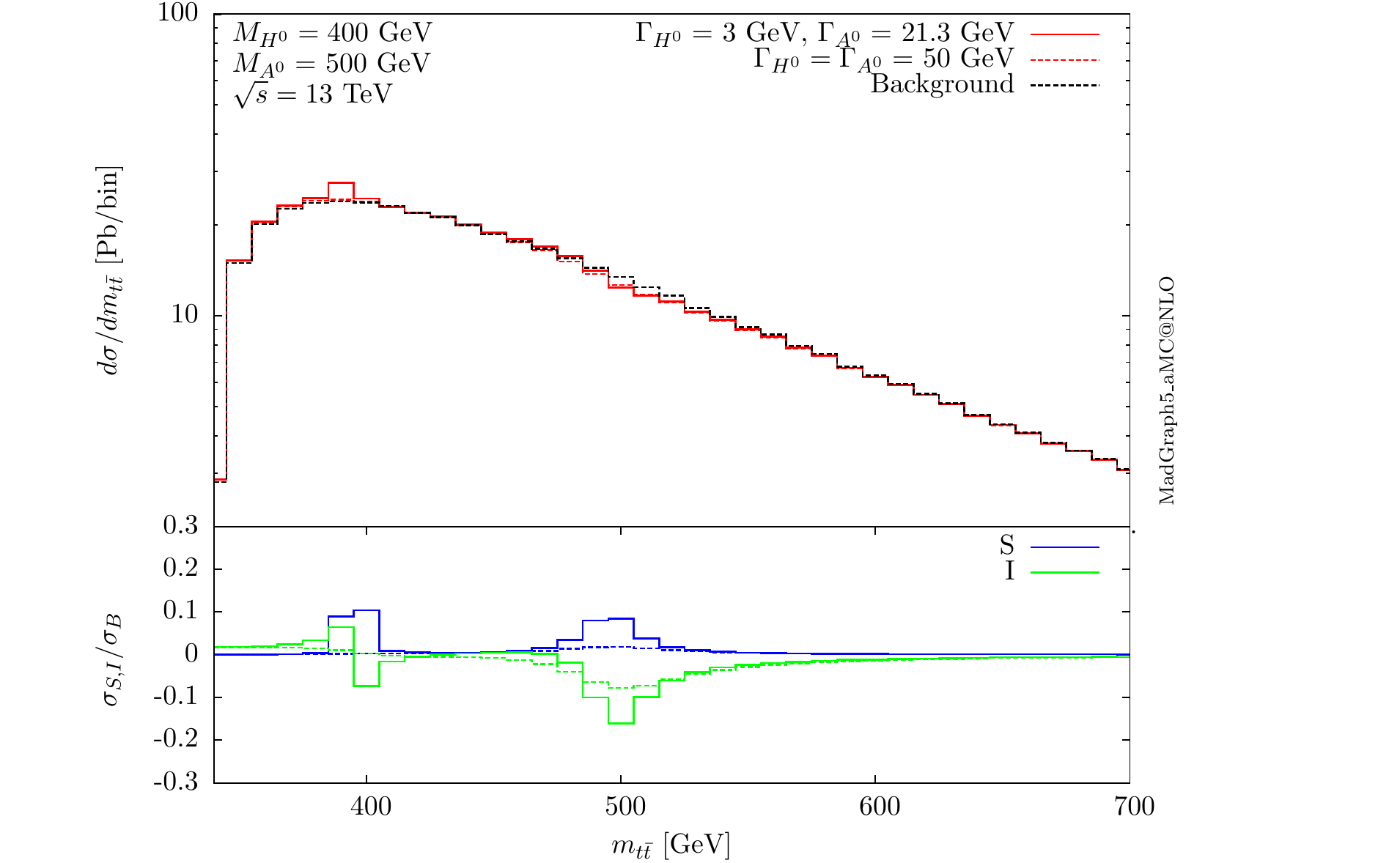}
\caption{Top pair invariant mass distribution for the LHC at 13 TeV for the $\Hzero$ and $\Azero$ resonances with different  masses and widths. Top left: small mass splitting ($\Delta m=10$ GeV), top right: moderate mass splitting ($\Delta m=30$ GeV), bottom left: $\Delta m=50$ GeV and bottom right: large mass splitting ($\Delta m=100$ GeV). The ratio of the signal and interference over the QCD background is shown in the lower panels.}
\label{LO_all}
\end{figure}

In figure~\ref{LO_all} we show the invariant mass distribution for a scenario in which both a scalar and pseudoscalar resonance are present.  The corresponding LO cross sections for the signal and interference with the QCD continuum are reported in table \ref{tbl:all}. In this case the mass splitting and widths of the two states are varied. The behaviour of the amplitude squared is replicated here. In the narrow width scenario, for a small mass splitting, we cannot distinguish between the two contributions. For $\Delta m=50$ GeV two separate peaks appear. In general, we find that the interference is destructive and large compared to the signal, in particular when the widths are large, a case where deviations from the background are generally suppressed. For all mass combinations, the interference is comparable in size with the signal even for the narrow width choices and its impact on the line-shape is important. The interference can lead to shapes very different from the resonance peaks that one expects from the signal alone. \\

\begin{table}[h!]
\begin{center}
    \begin{tabular}{|c|c|c|}
        \hline
$m_{H^0}=400$ GeV, $m_{A^0}=410$ GeV & $\Gamma_{min}$ & $\Gamma = 50$ GeV  \\ \hline
%Total & 491.81 & 487.13  \\ \hline
Signal & 17.56 & 3.09 \\ 
\hline 
Interference & -11.85 & -10.69 \\ 
\hline
$m_{H^0}=400$ GeV, $m_{A^0}=430$ GeV & $\Gamma_{min}$ & $\Gamma = 50$ GeV  \\ \hline
%Total & 493.47 & 488.78 \\ \hline
Signal & 14.92 & 2.97 \\ 
\hline 
Interference & -9.60 & -8.91 \\ 
        \hline 
$m_{H^0}=400$ GeV, $m_{A^0}=450$ GeV & $\Gamma_{min}$ & $\Gamma = 50$ GeV  \\ \hline
%Total & 494.52 & 490.04 \\ \hline
Signal & 12.92 & 2.77 \\ 
\hline 
Interference & -7.88 & -7.46 \\ 
\hline
     $m_{H^0}=400$ GeV, $m_{A^0}=500$ GeV & $\Gamma_{min}$ & $\Gamma = 50$ GeV  \\ \hline
%Total & 495.63 & 492.01  \\ \hline
Signal & 9.68 & 2.22 \\ 
\hline 
Interference & -5.23 & -4.94 \\ 
\hline 
\end{tabular}
 \caption{Cross sections (in pb) for the LHC at 13 TeV for a new heavy scalar and pseudoscalar particle for different widths and masses. Yukawas are equal to the SM value.
\label{tbl:all}}
\end{center} 
\end{table}

\begin{figure}[h!]
\centering
\includegraphics[scale=0.80]{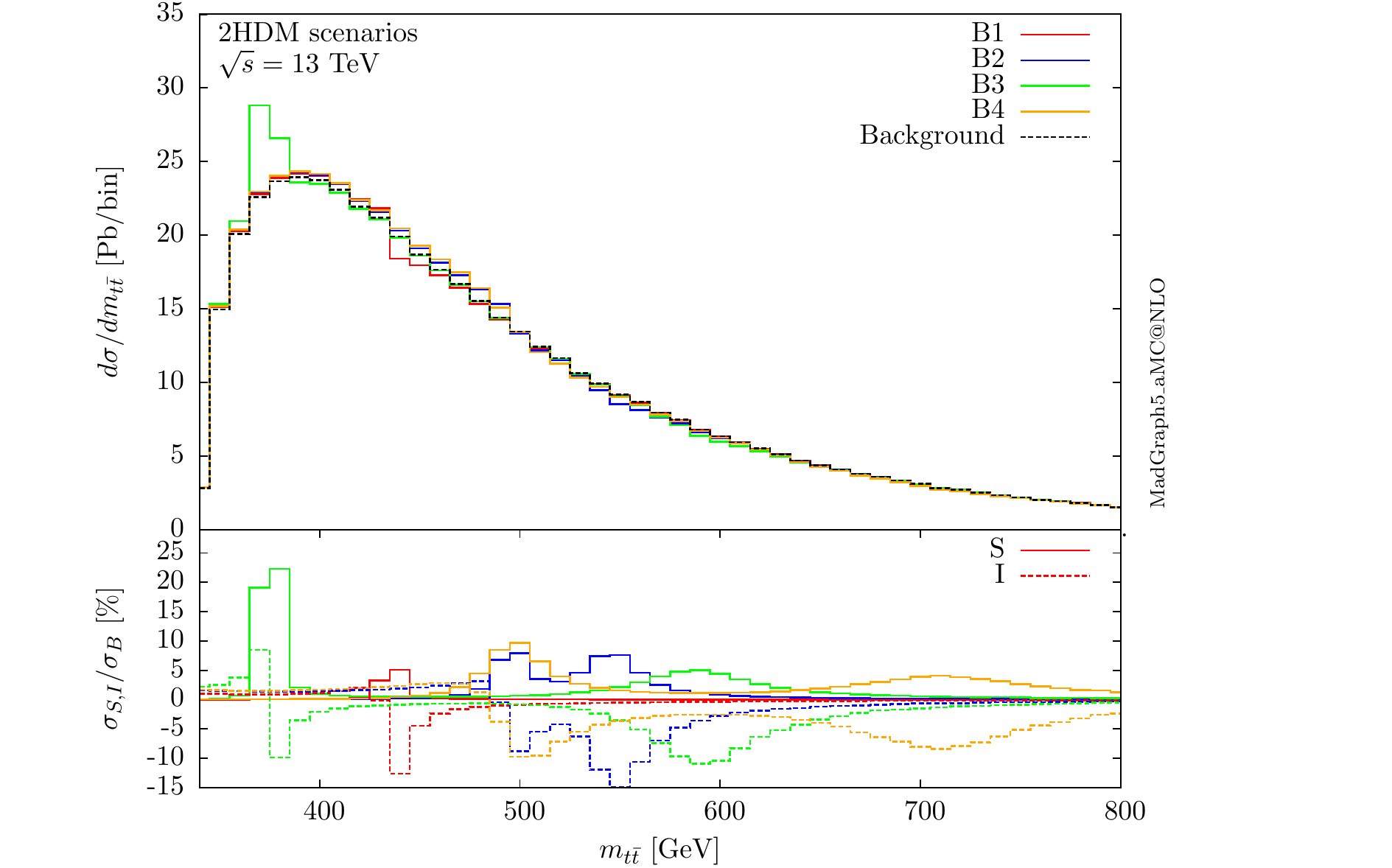}
\caption{Invariant mass distribution for the $t\bar{t}$ pair for the different 2HDM benchmark points. The ratios of the signal and interference over the QCD background are shown in the lower panel. }
\label{LO_2HDM}
\end{figure}

We conclude this section by considering the 2HDM benchmarks presented in section \ref{tab:benchmarks} to have a  picture of possible deviations from the SM predictions in a UV-complete model.  In this case the widths are computed using the 2HDM parameter input. The 2HDM parameters of interest i.e. the Yukawa couplings, the widths and top-quark branching ratios are given in table \ref{tab:couplings}. The corresponding cross sections are collected in table \ref{xsec2HDMLO}, where we also show the corresponding scale uncertainties obtained by varying the renormalisation and factorisation scales up and down by a factor of two. The interference is important and destructive for all scenarios, ranging in size from 40\% to 100\% of the signal at the total cross-section level. For completeness we also show the various contributions involving the light 125 GeV Higgs in table \ref{tab:xsec1}. These are found to be small in all cases. The differences between the four scenarios in the contributions involving only the light Higgs are due to the differences in the light Higgs Yukuwa coupling. The invariant mass distribution of the top quark pair for the four scenarios is shown in figure \ref{LO_2HDM}. \\

\begin{table}[h!]
\begin{center}
    \begin{tabular}{|c|c|c|c|c|c|}
        \hline 
        \multirow{2}{*}{Benchmarks}  & \multirow{2}{*}{Total} & \multicolumn{2}{|c|}{Signal} & \multicolumn{2}{|c|}{Interference}   \\ \cline{3-6}
        & & Scalar & Pseudoscalar & Scalar & Pseudoscalar \\ \hline
B1 & 497.05$^{+31.7\%}_{-22.7\%}$ & 0.01$^{+32.7\%}_{-23.1\%}$ & 2.13$^{+32.2\%}_{-22.8\%}$ & -0.62$^{+32.4\%}_{-23.0\%}$ & -1.78$^{+33.7\%}_{-23.6\%}$   \\ \hline
B2 & 501.01$^{+31.8\%}_{-22.7\%}$ & 2.90$^{+33.0\%}_{-23.3\%}$  & 3.51$^{+33.5\%}_{-23.5\%}$  & -1.55$^{+34.9\%}_{-24.3\%}$ & -1.09$^{+40.6\%}_{-27.1\%}$  \\ \hline
B3 & 503.96$^{+32.3\%}_{-23.2\%}$ & 10.86$^{+31.3\%}_{-22.4\%}$  & 3.35$^{+33.8\%}_{-23.7\%}$  & -6.56$^{+32.2\%}_{-22.9\%}$ & -1.15$^{+42.4\%}_{-28.0\%}$  \\ \hline
B4 & 502.06$^{+31.9\%}_{-22.8\%}$ & 6.19$^{+33.0\%}_{-23.3\%}$  & 1.85$^{+34.6\%}_{-24.1\%}$  & -3.53$^{+34.7\%}_{-24.1\%}$ & 0.30$^{+67.6\%}_{-56.0\%}$ \\ \hline
\end{tabular}
 \caption{Cross section at LO (in pb) for the LHC at 13 TeV for the 2HDM scenarios of table \ref{tab:benchmarks} with scale uncertainties. The signal and interference is decomposed into contributions from the scalar and pseudoscalar resonances.  \label{xsec2HDMLO} }

 %The Total corresponds to $pp \to t \bar{t}$, $ gg \to h \to t \bar{t}$, h-QCD interference, H-QCD and A-QCD interference, h-H interference and $gg \to H,A \to t \bar{t}$. 
%Signal is $gg \to H,A \to t \bar{t}$ and interference is h-H, H-QCD and A-QCD interference.  
\end{center} 
\end{table}
\begin{table}[h!]
\begin{center}
    \begin{tabular}{|c|c|c|c|}
        \hline
Benchmark  & $pp \to h \to t\bar{t}$ & h-H interference & h-QCD interference    \\ \hline
B1 & $0.019^{+33.2\%}_{-23.4\%}$ & $0.03^{+32.7\%}_{-23.4\%}$  &  $-0.82^{+32.1\%}_{-23.1\%}$ \\ \hline
B2 & $0.022^{+33.3\%}_{-23.5\%}$ & $0.01^{+34.2\%}_{-23.9\%}$ &  $-0.90^{+32.4\%}_{-23.2\%}$  \\ \hline
B3 & $0.022^{+33.3\%}_{-23.5\%}$ & $0.25^{+32.3\%}_{-23.0\%}$ &  $-0.90^{+32.4\%}_{-23.2\%}$  \\ \hline
B4 & $0.022^{+33.3\%}_{-23.5\%}$ & $0.03^{+34.3\%}_{-24.0\%}$ &  $-0.90^{+32.4\%}_{-23.2\%}$  \\ \hline
\end{tabular}
 \caption{Cross sections at LO (in pb) for the LHC at 13 TeV for the 2HDM scenarios of table \ref{tab:benchmarks} with scale uncertainties for the various contributions involving the light SM-like Higgs.   \label{tab:xsec1}}
\end{center} 
\end{table}

We find that benchmark B1 only shows deviations from the background around the mass of $\Azero$, as $\Hzero$ lies below the top--anti-top threshold.  The $\Azero$ contribution is dominated by the interference leading to a dip in the invariant mass distribution. Scenario B2 shows a more involved structure due to the presence of two resonances with a 50 GeV mass splitting. All Yukawas are enhanced, nevertheless the large widths and the cancellation of the destructive interference with the signal lead to effects of a few percent compared to the background. Benchmark B3 is the only scenario that leads to a visible resonance peak at  380 GeV and a mild dip at around 590 GeV, corresponding to the narrow $\Hzero$ and broad $\Azero$ resonances respectively. Finally B4 shows a small excess over the background at 500~GeV and a mild dip at around 700 GeV. Due to the large widths the effects on the invariant mass distribution are extremely mild and would therefore be difficult to detect.

\section{Higher-order QCD effects}
\label{higherorder}
\subsection{Signal-background interference in $t\bar{t}+$jet}
As we have seen in the previous subsection the interference between the signal and background can lead to interesting peak-dip structures, and needs to be taken into account to obtain a reliable prediction for the line-shape of a new resonance. It is well-known that the interference between the signal and the QCD background is colour-suppressed: i.e. the only background configuration which contributes to the interference is the one where the top--anti-top pair is in a colour singlet state. In this section we investigate whether this colour suppression could be lifted by allowing additional QCD radiation.  

\begin{figure}[h!]
\centering
\includegraphics[scale=0.45, trim=3cm 5cm 2cm 0]{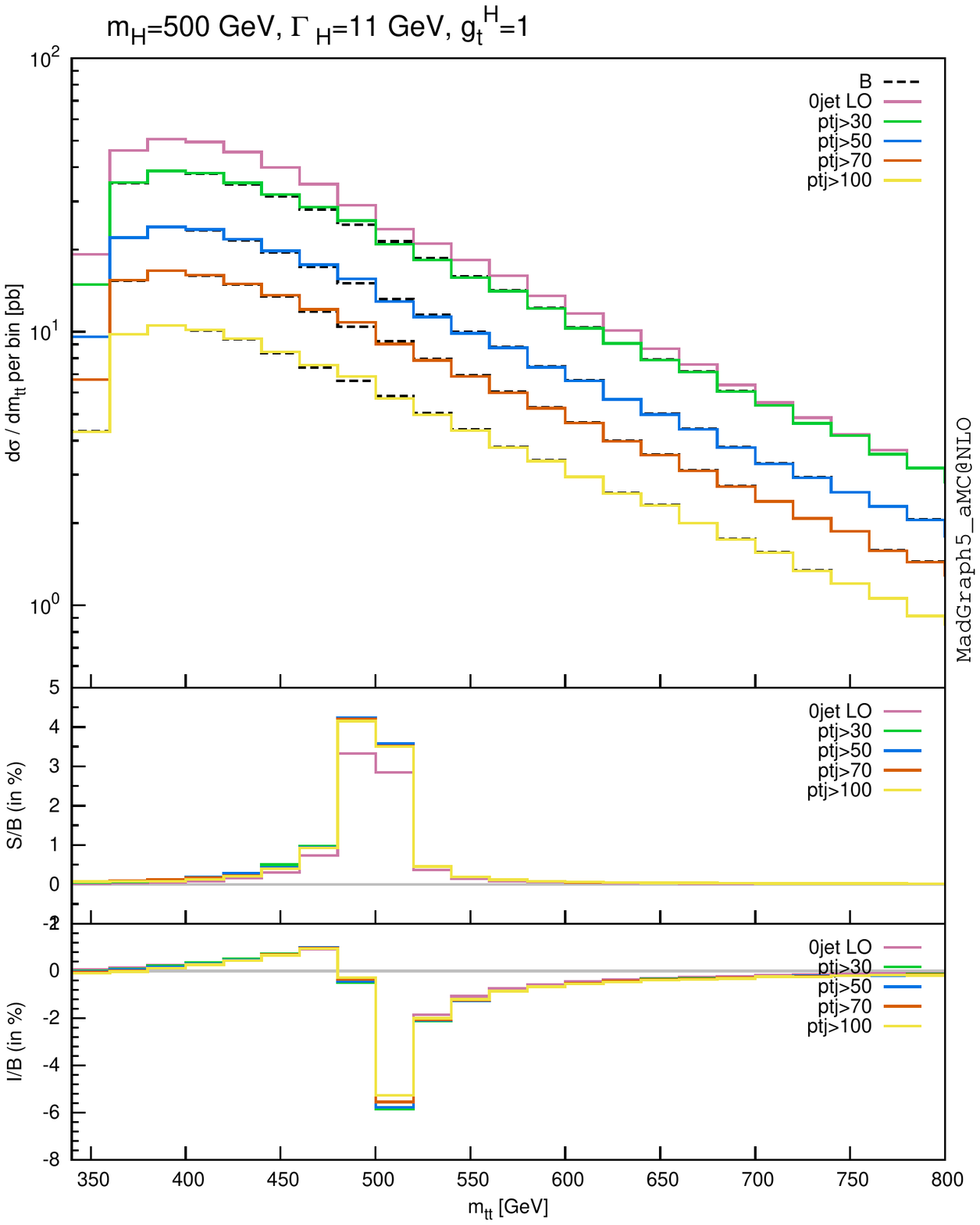}
\includegraphics[scale=0.45, trim=2cm 5cm 3cm 0]{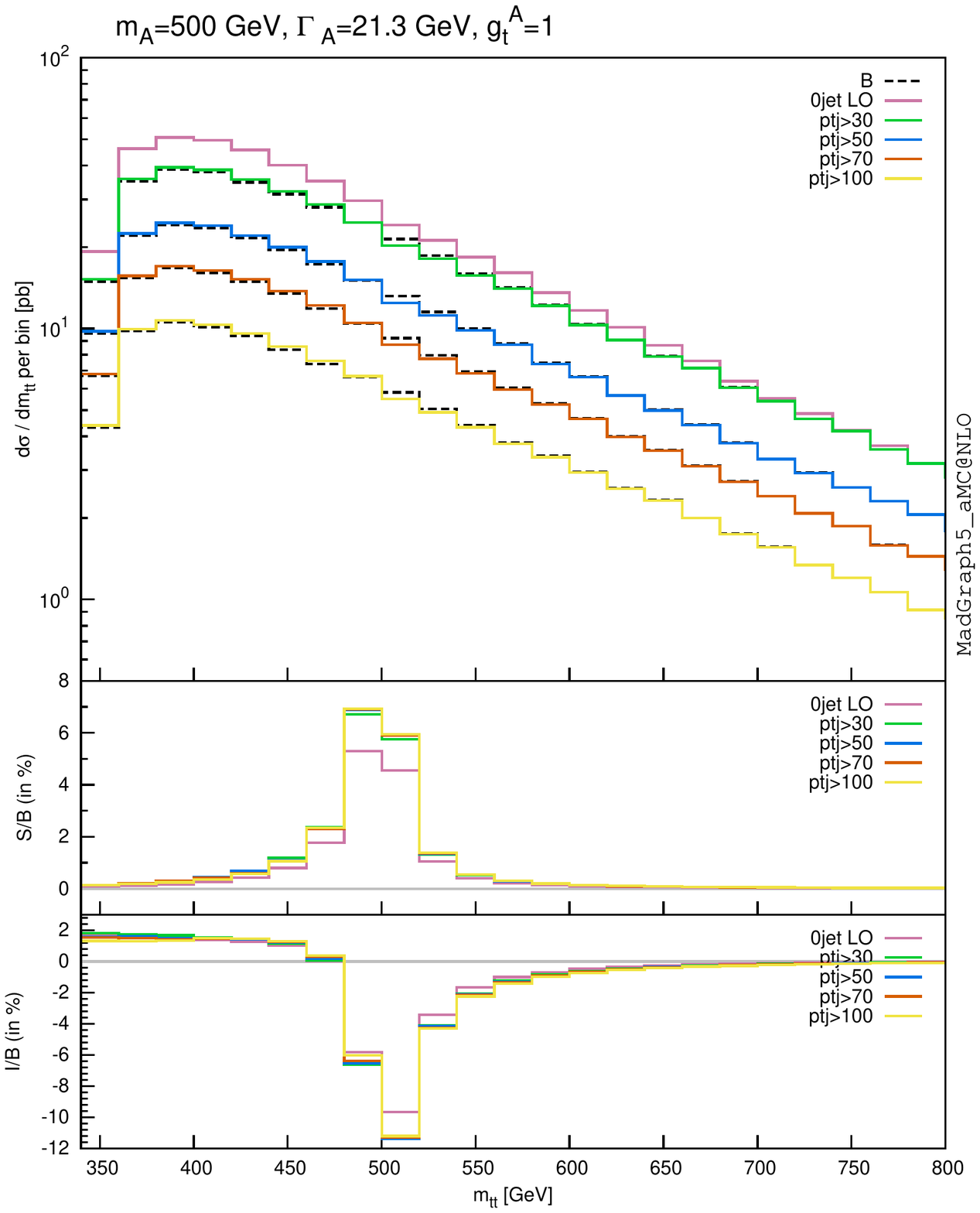}
\caption{Top pair invariant mass distribution for the LHC at 13 TeV for $p p(\to Y) \to  t\bar{t} (j)$  for a heavy scalar or pseudoscalar with $m_Y=500$ GeV. Different $p_T^j$ cuts are applied for the $t\bar{t} j$ process. The lower panels show the signal and interference ratios over the background. }
\label{1jet_SM500}
\end{figure}

We consider for the first time signal and background interference effects for the $t\bar{t}+$jet process. Figure~\ref{1jet_SM500} shows a comparison between the LO process $p p (\to Y) \to t\bar{t}$ and the one with an additional jet emission $p p (\to Y) \to  t\bar{t} j$, for a scalar or pseudoscalar of $m_{Y}=500$ GeV and $\Gamma_{Y}=\Gamma_{min}$.  For the $p p (\to Y) \to  t\bar{t} j$ process a cut has to be applied on the jet transverse momentum. The ratio of the signal and interference over the background is shown for various cuts on the jet $p_T$. We find that the extra jet does not give rise to a significant increase of the interference. For the scalar resonance the relative size of the interference is identical to that for the $2\to 2$ process, while for the pseudoscalar a small increase is found. The line-shape of the resonance is not visibly modified by the QCD radiation. This persists even for very hard jets for which the cross-section is as expected suppressed. We associate this to the fact that the main contribution to the 1-jet process is related to initial state radiation, for which no change in the colour state of the top-quark pair is expected and therefore the colour suppression is not lifted.

We note here that a consistent way to include both 0 and 1 jet multiplicities is to employ the Matrix-Element--Parton Shower (ME+PS) method.  Such a procedure is possible within {\sc MadGraph5\_aMC@NLO}. Merged samples can be passed to {\sc Pythia 6} or {\sc Pythia 8}  \cite{Sjostrand:2007gs,Sjostrand:2014zea} for PS. Given the results obtained for the 1-jet samples which show the very mild effect of the extra QCD radiation on the interference between signal and background and hence the line-shape of a heavy scalar, we refrain from performing a detailed analysis of a merged sample.

\subsection{NLO results}
\label{NLO}
\begin{figure}[h!]
\centering
\includegraphics[scale=0.8]{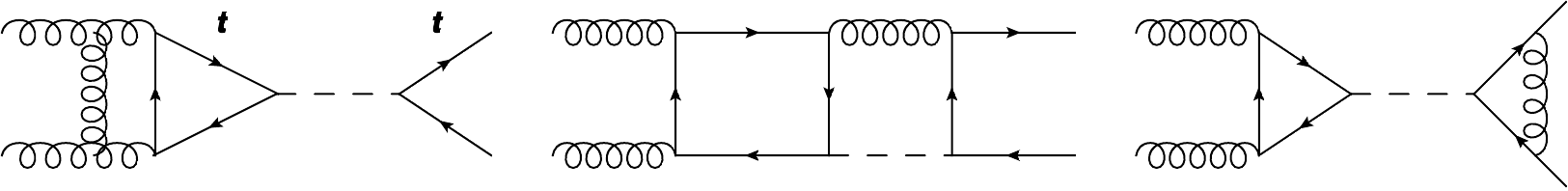}
\caption{Two-loop virtual corrections diagrams for the heavy scalar signal.}
\label{virtual}
\end{figure}

In order to improve the accuracy of the predictions for this process, we now examine the impact of NLO corrections. We start by reviewing the main ingredients needed for the computation of the signal, background and interference at NLO. The NLO QCD corrections for the signal require 1-loop real emission amplitudes and 2-loop virtual correction amplitudes. A sample of the required diagrams is shown in Figure \ref{virtual}. These can be classified in three categories: initial state corrections, final state corrections and corrections connecting initial and final state, the so-called non-factorisable corrections. The initial state corrections are identical to the NLO corrections for single Higgs production and are well known \cite{Spira:1995rr,Harlander:2005rq}. The final state corrections are also well known as part of the QCD corrections to the Higgs decay width to heavy quarks \cite{Djouadi:1995gt}. Results are not available for the class of two loop amplitudes shown in the centre of figure \ref{virtual}, as these require multiscale integrals at the edge of current multiloop technology.  Exact results can be obtained for the signal at NLO, as this class of diagrams does not interfere with the Born amplitude as in the Born configuration the top quark pair is in a colour singlet. The non-factorisable corrections only play a role in the interference between the signal and the continuum background, which is therefore formally known only at leading order. 

An approximation to the NLO results has been presented in \cite{Bernreuther:2015fts}, where two approximations are made. The first regards the interaction of the Higgs to the gluons, which is computed in the infinite top mass limit. The second is the computation of the NLO QCD corrections for the signal and interference in the soft gluon approximation. 
In this work we follow a different approach. We compute the NLO corrections for the signal with the exact top mass dependence, while for the interference we employ a $K$-factor obtained from the geometric average of the signal and background $K$-factors. The $K$-factor approximation can be employed both at the total cross-section level and on a bin-by-bin basis for the differential distributions. A similar procedure is recommended for other loop-induced processes such as $g g (\to H) \to VV$ which also suffer from the lack of two-loop results. In the context of this study we have explicitly verified that the geometric average of the signal and background ratios of 1-jet over 0-jet cross section provides a good approximation for the corresponding ratio for the interference in the proximity of the resonance mass. 

On the computational side, within {\sc MadGraph5\_aMC@NLO}, the background can be obtained automatically at NLO. For the signal the two loop virtual corrections for Higgs production are taken from those in SusHi \cite{Harlander:2012pb} as implemented in aMCSusHi \cite{Mantler:2015vba}. These are combined with the 1-loop corrections in the final state which are computed with {\sc MadLoop}. The full 1-loop real and born amplitudes and 2-loop virtual corrections are inserted in the computation through a reweighting procedure.   

We decompose the total cross section using the following additive prescription:
\begin{equation}
\sigma_{NLO}=\sigma^{back}_{NLO}+\sigma^{signal}_{NLO}+\sigma^{inter}_{LO} \sqrt{K_S K_B}\,,
\end{equation}
where the signal and background are computed exactly at NLO in QCD. $\sqrt{K_S K_B}$ can involve either the total cross-section $K$-factors for the signal and the background or the bin-by-bin $K$-factors in the invariant mass spectrum as well as for any other observable of interest.

\begin{figure}[h]
\centering
\includegraphics[scale=0.42, trim=4cm 6cm 3cm 2cm]{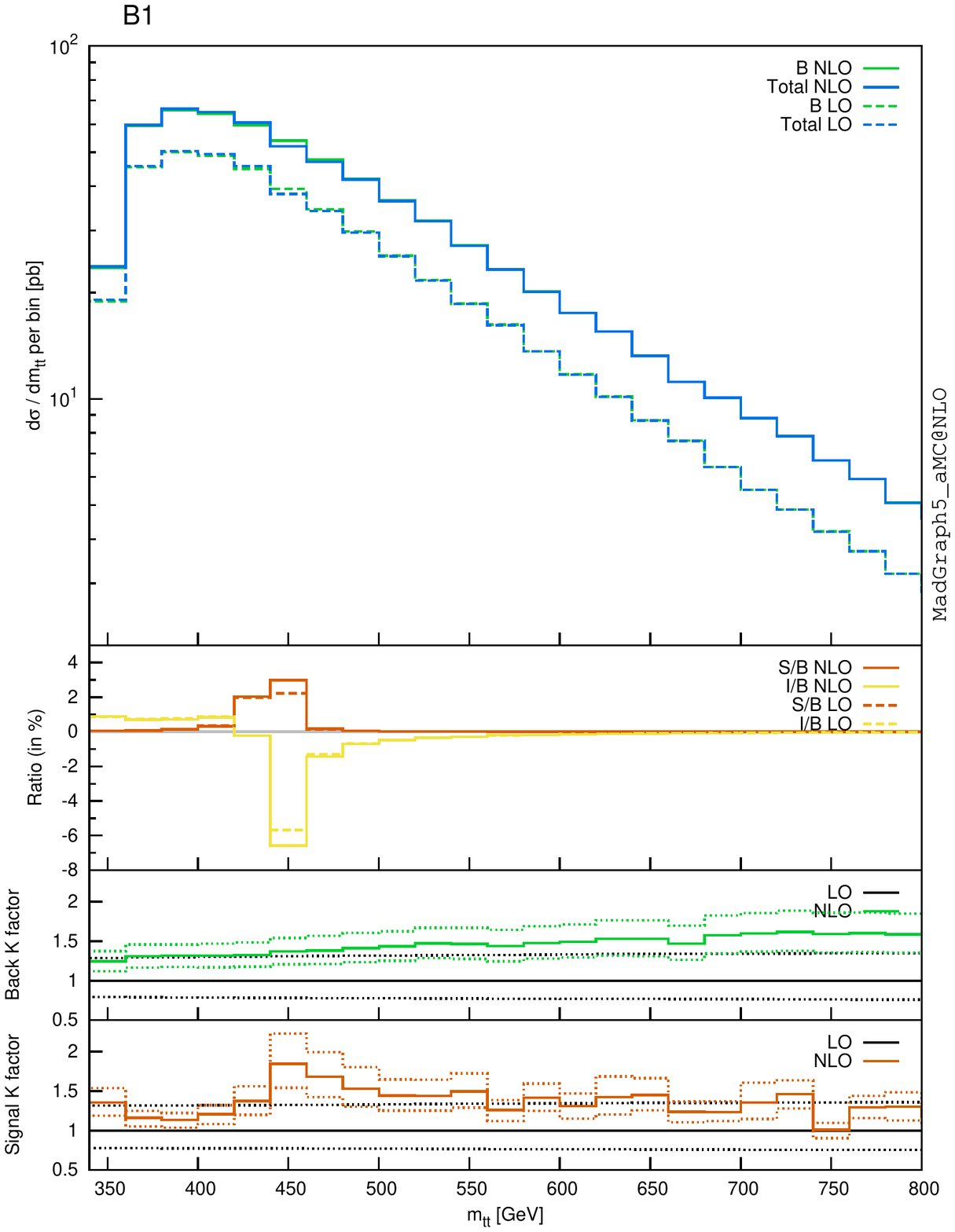}
\includegraphics[scale=0.42, trim=2cm 6cm 3cm 2cm]{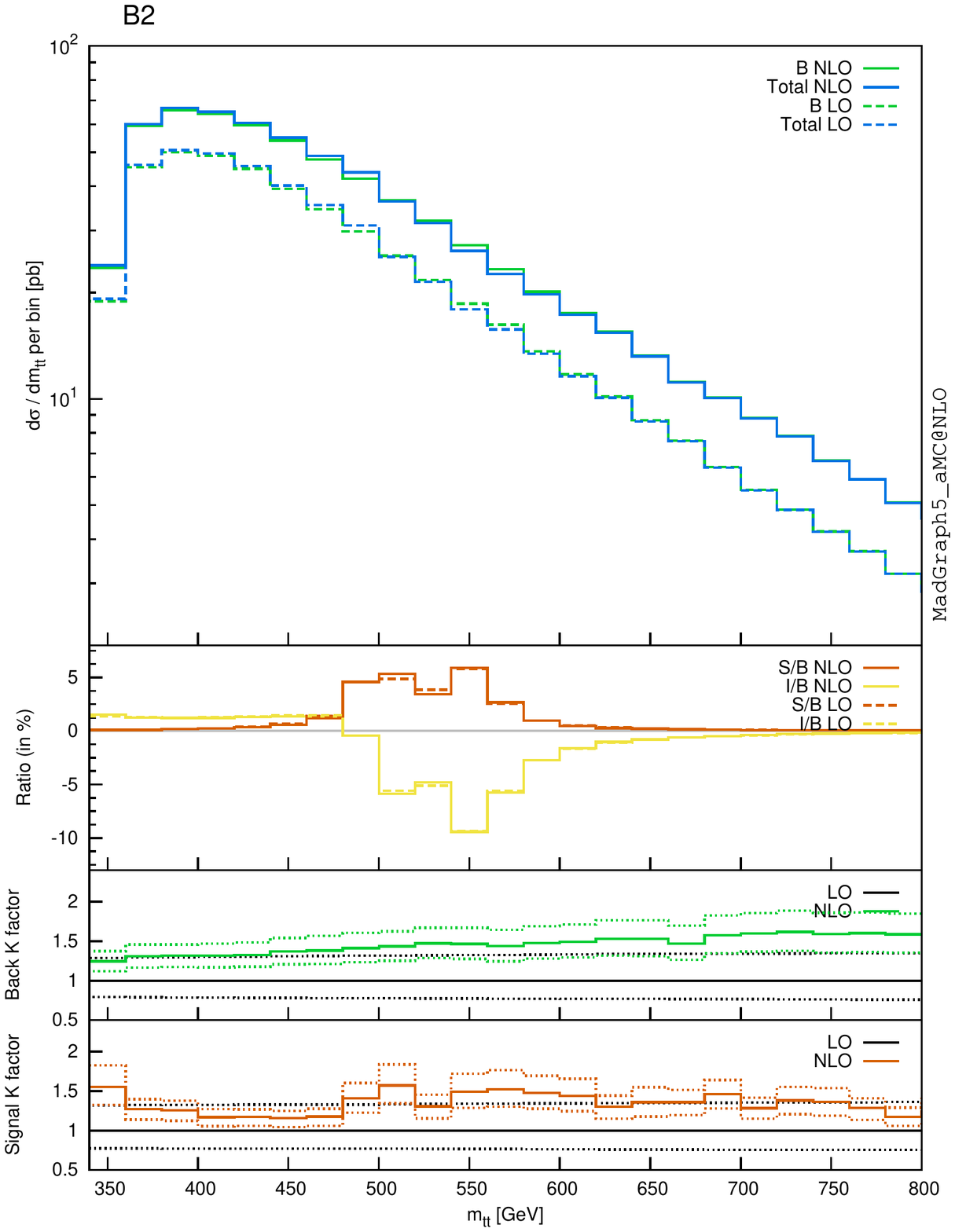}
\includegraphics[scale=0.42, trim=4cm 5cm 3cm 0]{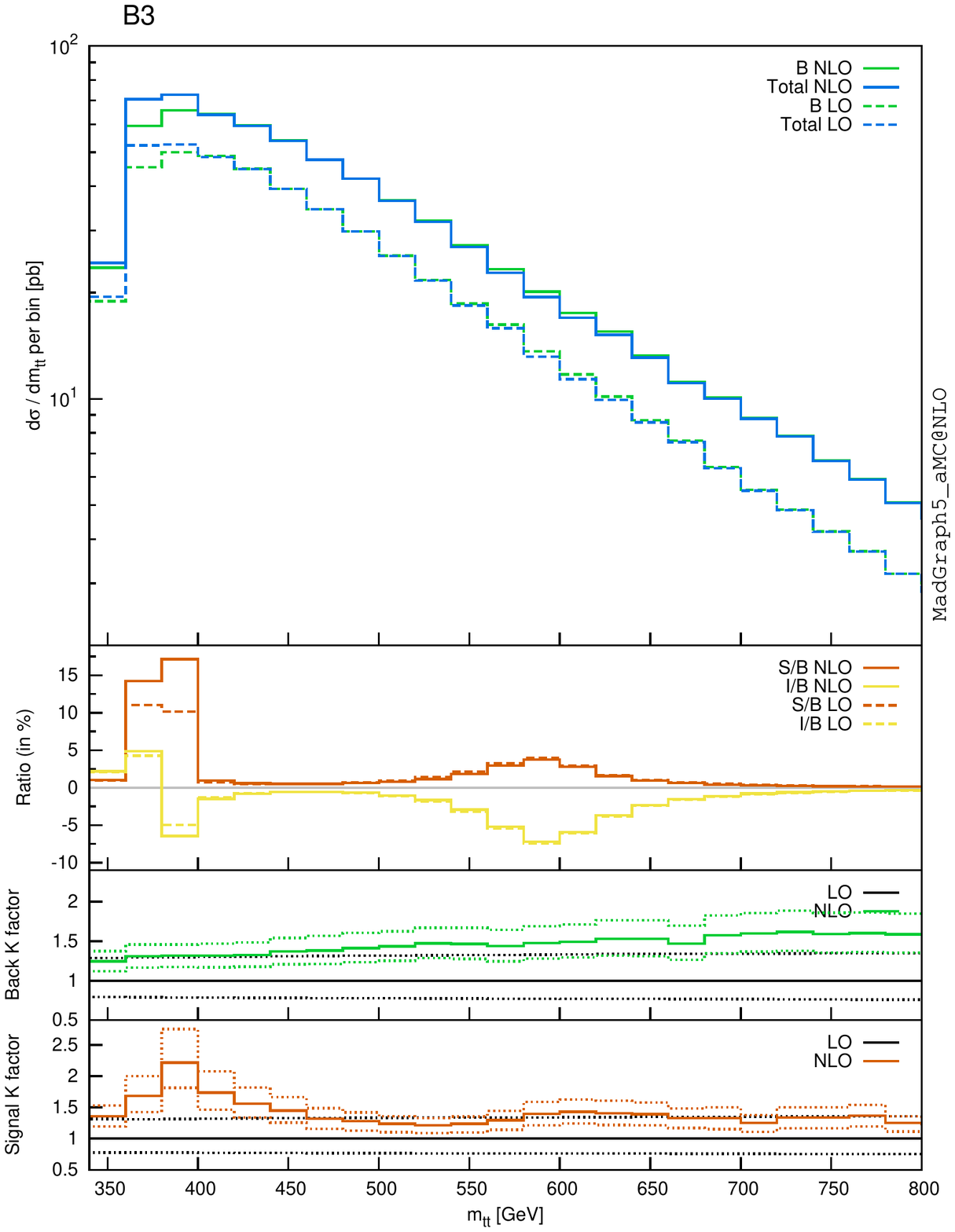}
\includegraphics[scale=0.42, trim=2cm 5cm 3cm 0]{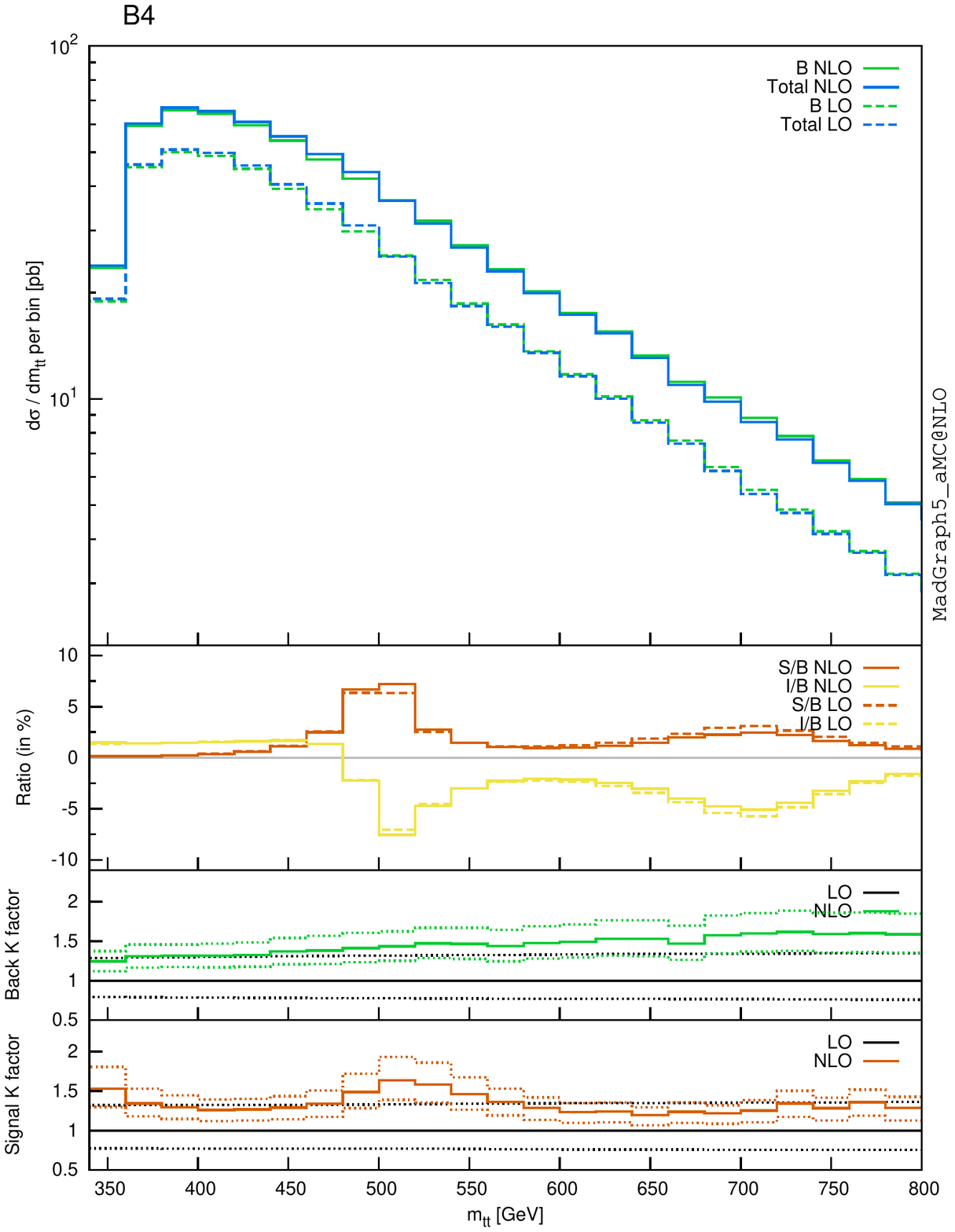}
\caption{Top pair invariant mass distribution for the different 2HDM benchmark points at NLO for the LHC at 13 TeV. The signal and interference ratios over the background are shown in the second panel, while the third and fourth panels show the background and signal $K$-factors along with the corresponding scale uncertainty bands. }
\label{NLO_HH}
\end{figure}

For brevity we present results at NLO only for our four 2HDM benchmarks. Results for the simplified model can be straightforwardly obtained with our setup. In table \ref{table:xsec2HDMNLO} the signal at NLO with the scale uncertainties, the corresponding $K$-factors and the NLO approximation for the interference are given for the four scenarios. The interference is computed at LO with NLO PDFs and the result is subsequently adjusted by the $K$-factor. The total cross-section $K$-factors are used to obtain the interference $K$-factor used in table \ref{table:xsec2HDMNLO}. We note that the scale uncertainties for the interference are those obtained from a LO computation and therefore are much larger than those of the signal and background. For the interference, our results provide a more accurate prediction, however we do not improve the precision of this contribution and therefore keep the LO uncertainties. For completeness we mention the NLO QCD background cross section $\sigma_{QCD}=  698.6^{+13.2\%}_{-12.4\%}$ pb and the corresponding $K$-factor $K_B=1.40$.  

The top pair invariant mass distribution for the LHC at 13 TeV is shown in figure \ref{NLO_HH}. The ratios of the signal and interference over the background are shown at LO and NLO, along with the signal and background $K$-factors with the corresponding scale uncertainties. We find large QCD corrections for the signal, with $K$-factors reaching two close to the resonance. The background $K$-factor is lower but rises with $m_{t\bar{t}}$.  Due to the larger $K$-factor for the signal compared to the background we notice an increase of the signal and interference over background ratios. The significant reduction of the scale uncertainties at NLO is also evident in the results. We note here that for the distributions we have extracted the $K$-factor for the interference using the signal and background $K$-factors in each bin.

\begin{table}[h!]
\begin{center}
    \begin{tabular}{|c|c|c|c|}
        \hline 
Benchmark   & Signal  & $K_S$& Interference: $\sqrt{K_S K_B}\,\sigma_{LO}^{inter}$    \\ \hline
B1  & $3.31^{+16.8\%}_{-14.3}$& 1.55  & -3.00$^{+30.7\%}_{-22.4\%}$ \\ \hline
B2  & $9.02^{+13.7\%}_{-13.0}$ & 1.41& -3.53$^{+34.3\%}_{-24.1\%}$   \\ \hline
B3 & $25.37^{+20.0\%}_{-16.0}$ & 1.79 & -9.32$^{+32.6\%}_{-24.1\%}$ \\ \hline
B4  & $11.51^{+14.3\%}_{-13.3\%}$ & 1.43 & -4.23$^{+38.8\%}_{-30.1\%}$ \\ \hline 
\end{tabular}
 \caption{Cross sections and corresponding scale uncertainties at NLO (in pb) for the LHC at 13 TeV for the 2HDM scenarios. The corresponding $K$-factors and the interference with the QCD background obtained from the geometric average of the signal and background total cross-section $K$-factors ($K_B=1.40$) are also given.\label{table:xsec2HDMNLO}}
\end{center} 
\end{table}

\section{Comparison with experimental measurements}
\label{constraints}
Our improved theoretical predictions can be used along with the experimental measurements of the top pair production cross section to obtain constraints on new physics contributions. In particular we employ the ATLAS $t\bar{t}$ resonant search \cite{Aad:2015fna} to set constraints on a simplified model with an extra scalar or pseudoscalar particle coupling to the top quark. Our results can be reinterpreted in terms of 2HDM scenarios and be combined with other constraints. 

The ATLAS 8~TeV $t\bar{t}$ resonance search~\cite{Aad:2015fna} uses the reconstructed invariant mass of the top quark pair, to place 95\% C.L. exclusion on the existence of scalar resonances coupling to top quarks. The search sets limits on the resonant cross section ranging from 3.0~pb for a mass of 400~GeV to 0.03~pb for 2.5~TeV. The results assume a narrow width approximation, {\it i.e.}  a total width $\lesssim 3\%$ of the mediator mass. 

\begin{figure}[h!]
\centering
\includegraphics[scale=0.4]{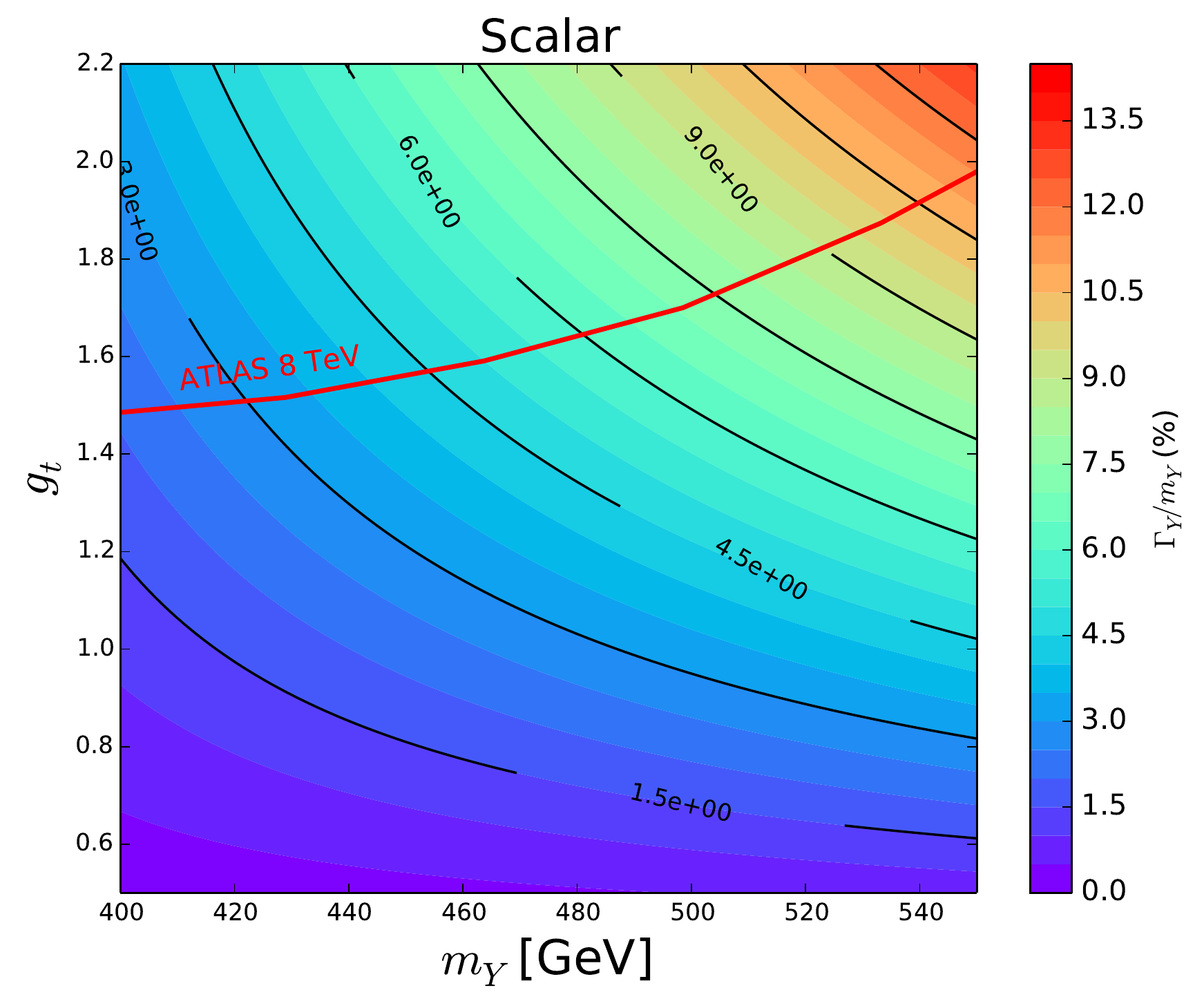}
\includegraphics[scale=0.4]{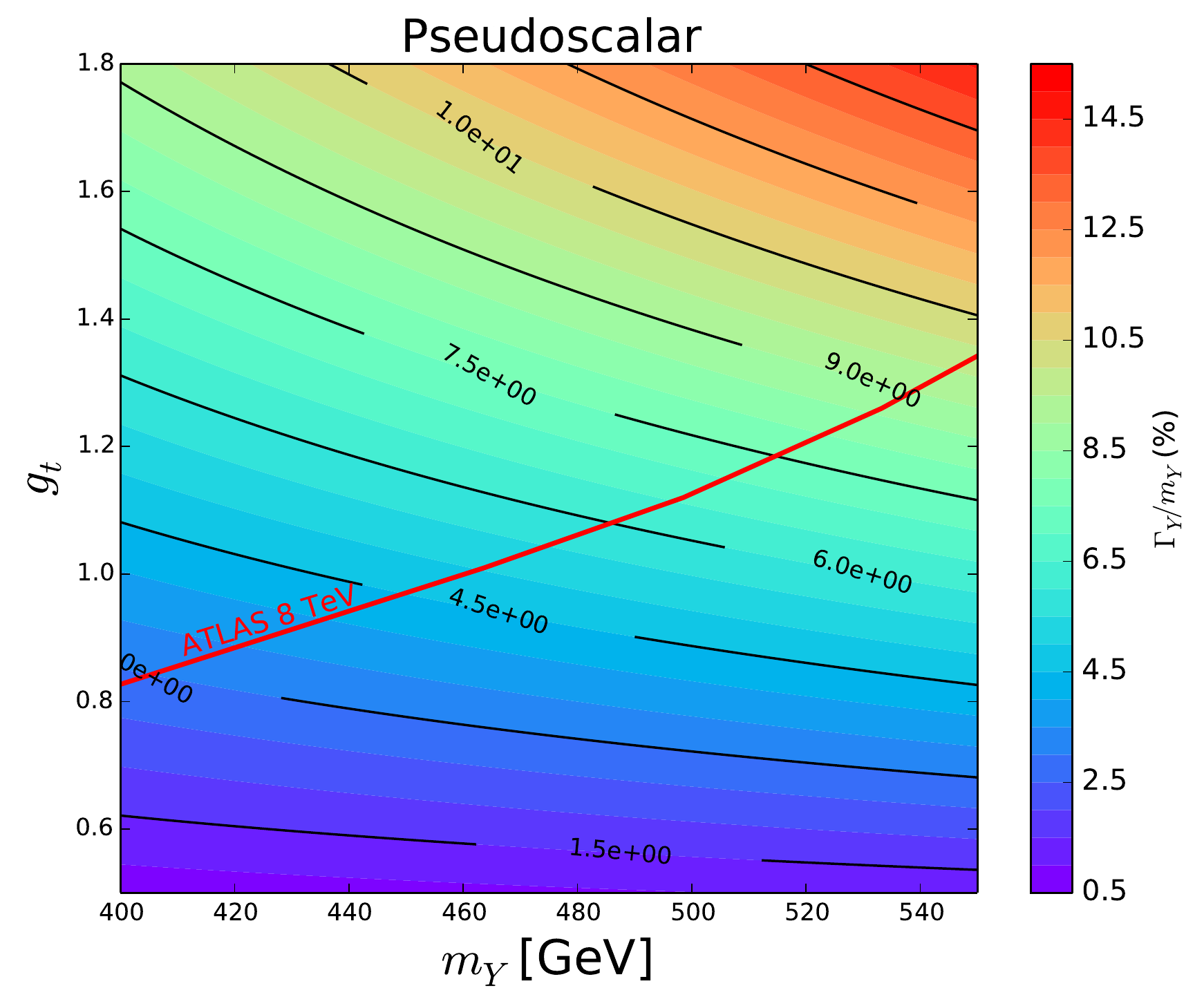}
\caption{Width of a spin-0 resonance as a percentage of its mass as a function of the mass and coupling to the top. Left: scalar. Right: pseudoscalar. The region above the red curve labeled ``ATLAS 8~TeV'' is excluded when LO predictions for the signal are used. }
\label{fig:topATLAS}
\end{figure}

Figure~\ref{fig:topATLAS} shows the width of the resonance coupling to the top only as a function of its mass and the top Yukawa coupling for a scalar and pseudoscalar. The ATLAS 8~TeV $t\bar{t}$ resonance search limit is also shown on the plot, extracted by converting the 95\% C.L. cross section to a value of the coupling using LO predictions. The region above the red line labelled ``ATLAS 8~TeV'' is excluded, if one assumes that the scalar particle only couples to the top quark. Our results show that scalar mediators with masses from 400~GeV to 550~GeV could be excluded for couplings $g_t\gtrsim 1.5$ depending on the mass of the mediator. For a pseudoscalar smaller values of the coupling can be excluded as the production cross-section is larger for a pseudoscalar resonance. While the search extends to much larger masses of mediators we do not show any results above 550 GeV as within this model it is not feasible to obtain a limit satisfying the narrow width approximation. As shown in figure ~\ref{fig:topATLAS}, the width over mass ratio rises quickly with $g_t$ and $m_Y$. In order to apply the ATLAS results we allow widths below 8\%  of the mass (which is the experimental resolution of the invariant mass of the $t \overline{t}$ system), which allows masses up to 550~GeV to be tested.

\begin{figure}[h!]
\centering
\includegraphics[scale=0.4]{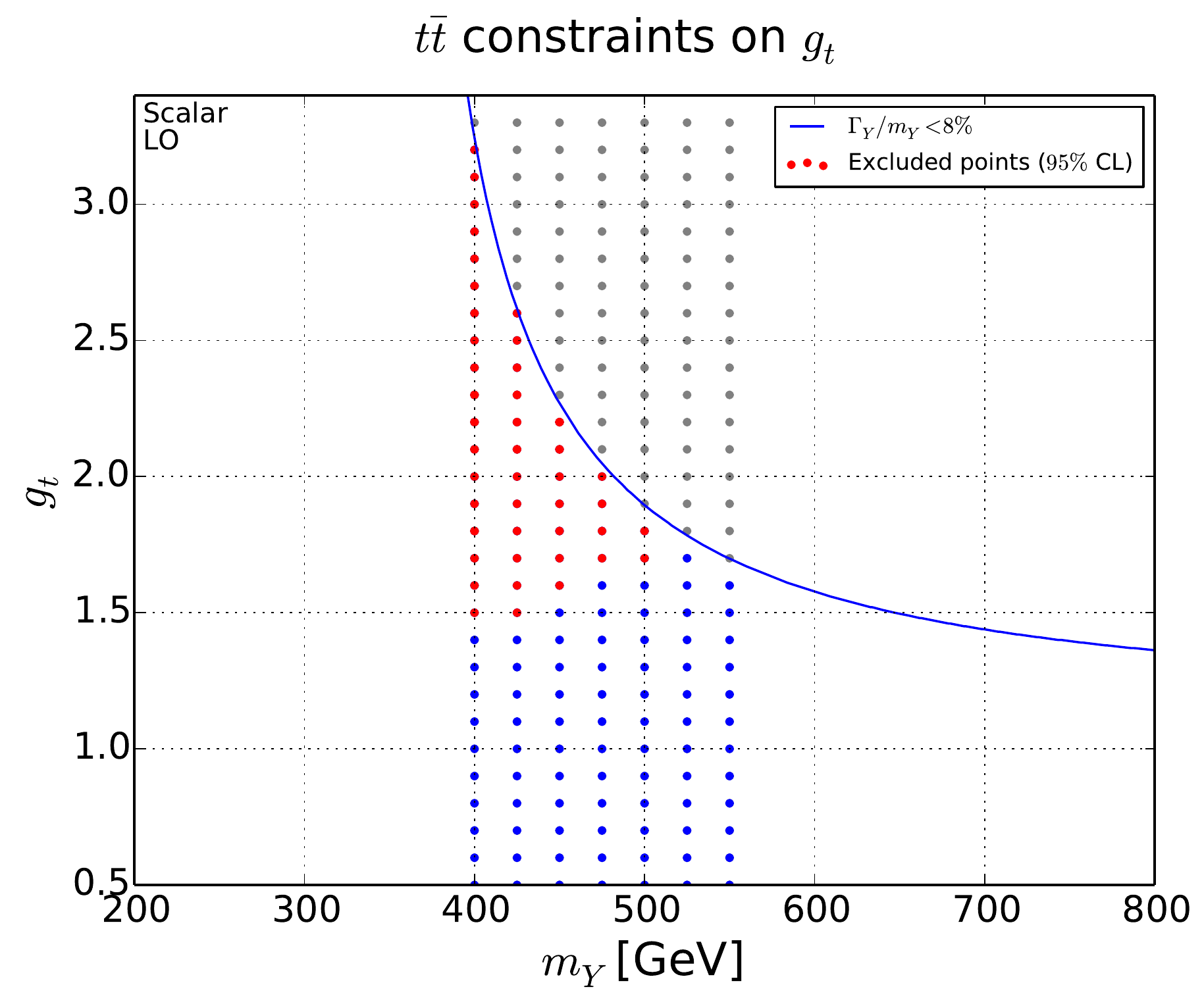}
\includegraphics[scale=0.4]{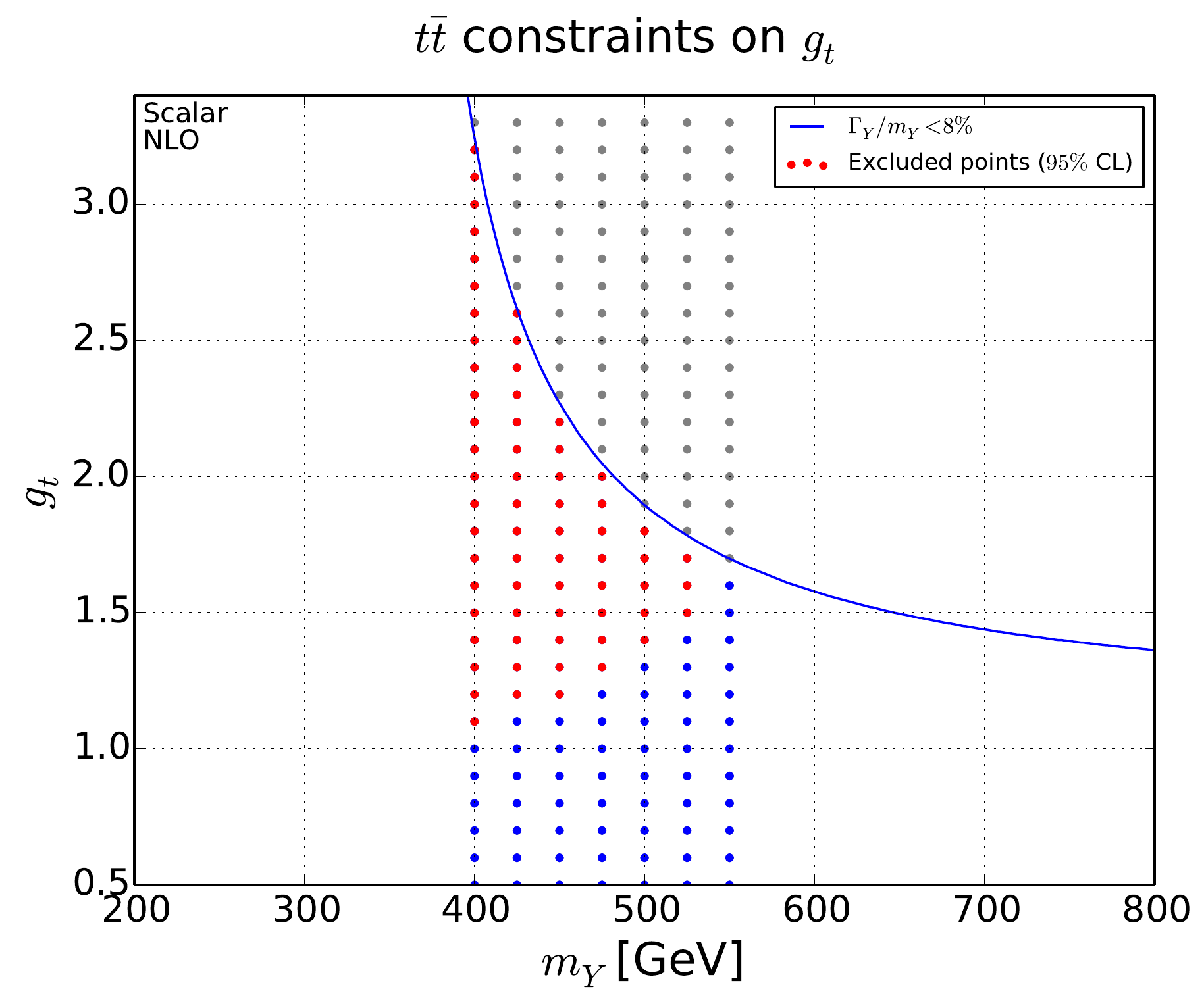}
\caption{Exclusion region obtained using the ATLAS 8~TeV $t\bar{t}$ resonance search results~\cite{Aad:2015fna} for a scalar resonance coupling to the top only using LO (left) and NLO (right) predictions for the signal cross section. }
\label{fig:exclusion}
\end{figure}

\begin{figure}[h!]
\centering
\includegraphics[scale=0.4]{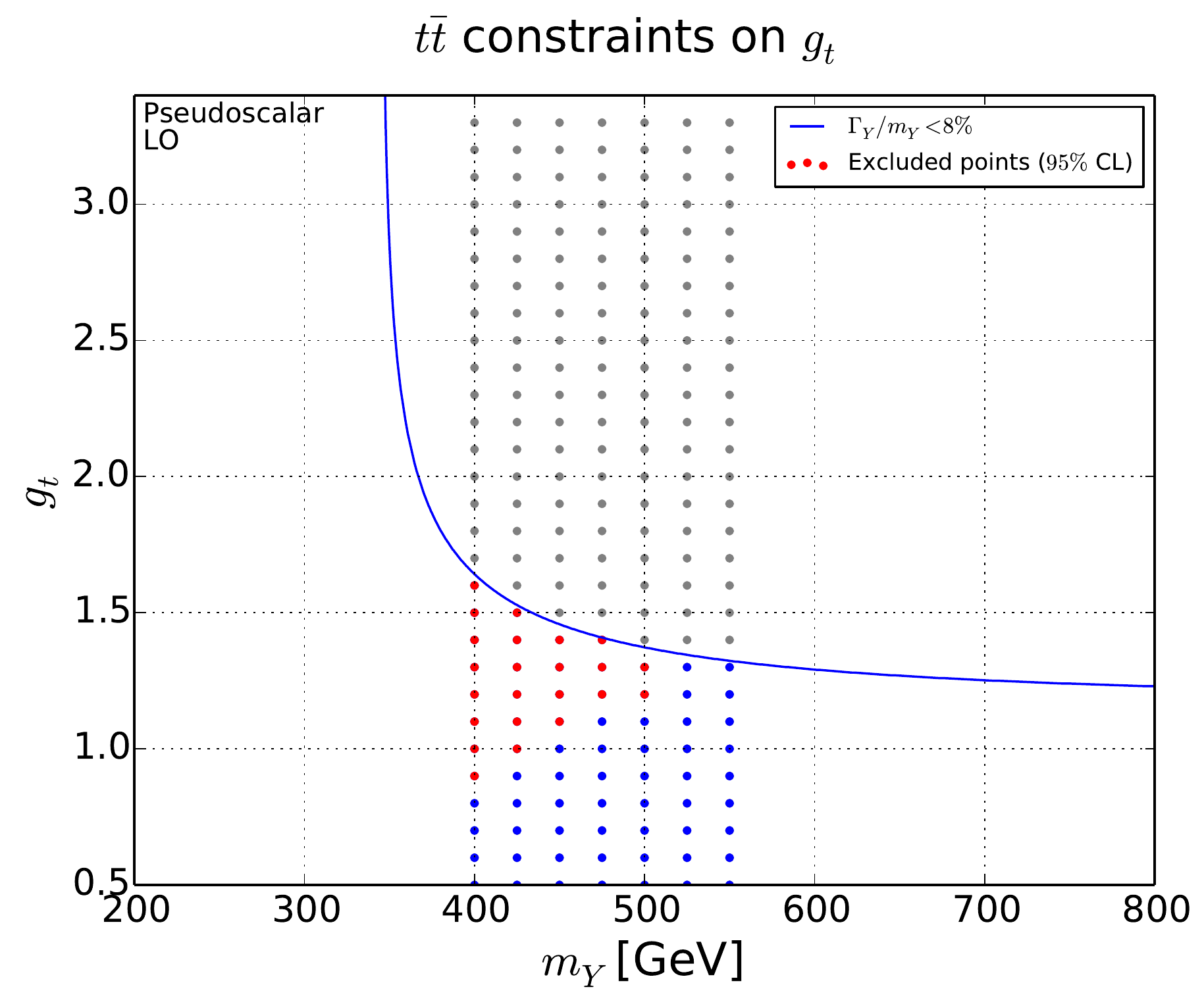}
\includegraphics[scale=0.4]{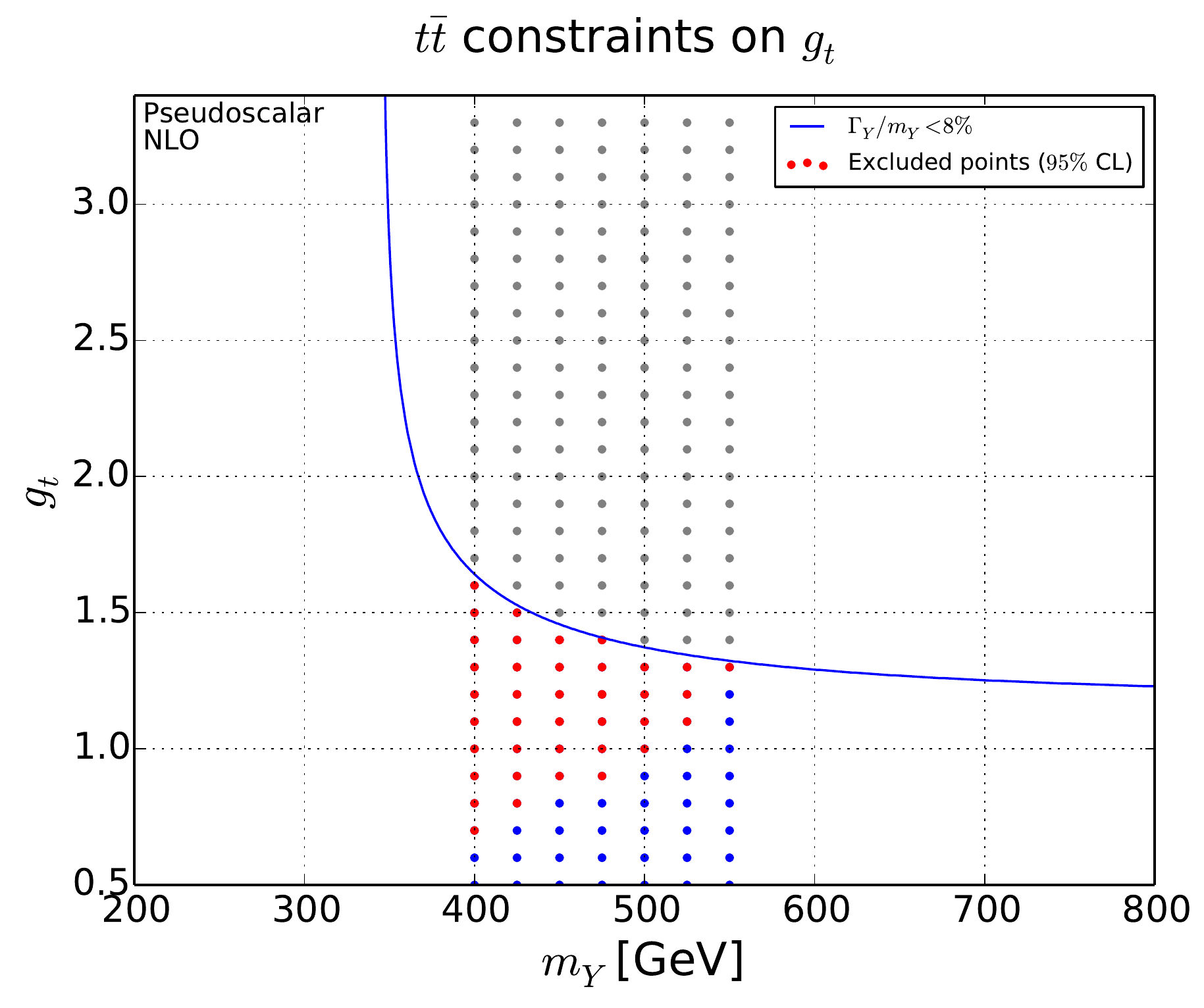}
\caption{Exclusion region obtained using the ATLAS 8~TeV $t\bar{t}$ resonance search results~\cite{Aad:2015fna} for a pseudoscalar resonance coupling to the top only using LO (left) and NLO (right) predictions for the signal cross section.}
\label{fig:exclusionpseudo}
\end{figure}

Focussing on this region, as shown in figure \ref{fig:topATLAS} for both scalar and pseudoscalar resonances, the width remains small for couplings $g_t<2$. Using our signal predictions at LO and NLO we extract the exclusion region in figure \ref{fig:exclusion} and \ref{fig:exclusionpseudo} for a scalar and pseudoscalar resonance.  As expected the exclusion region extends to smaller values of the coupling when we use the NLO predictions. We find that for the scalar mediator a larger region is excluded compared to the pseudoscalar one. The reason is the fact that the narrow width approximation is valid for larger values of the coupling for a scalar mediator compared to the pseudoscalar one. This allows us to apply the ATLAS results for a wider range of couplings for the scalar mediator.

We note here that the interference between signal and background is not taken into account by the ATLAS analysis. This implies that the search is based on the assumption that the signal will appear as a Breit-Wigner resonance over the SM background. In order to allow for the interference to be taken properly into account the experimental strategy would have to be modified, as the interference can lead to shapes which are very different from those expected for the signal only. This is particularly important when searches start focussing on resonances which are not extremely narrow. As we have already seen, the larger the width the bigger the impact of the interference. 
It is clearly not possible to a posteriori account for the impact of any potential shape changes, i.e. deviations from a Breit-Wigner resonance shape, on the 95\% C.L. exclusion cross-section obtained by ATLAS. Nevertheless we can estimate how including the interference at the total cross-section level can modify the limits set on the coupling in our simplified model, in the cases where the interference does not completely dominate the BSM contribution and therefore the shape of the deviation from the background. 

\begin{figure}[h!]
\centering
\includegraphics[scale=0.4]{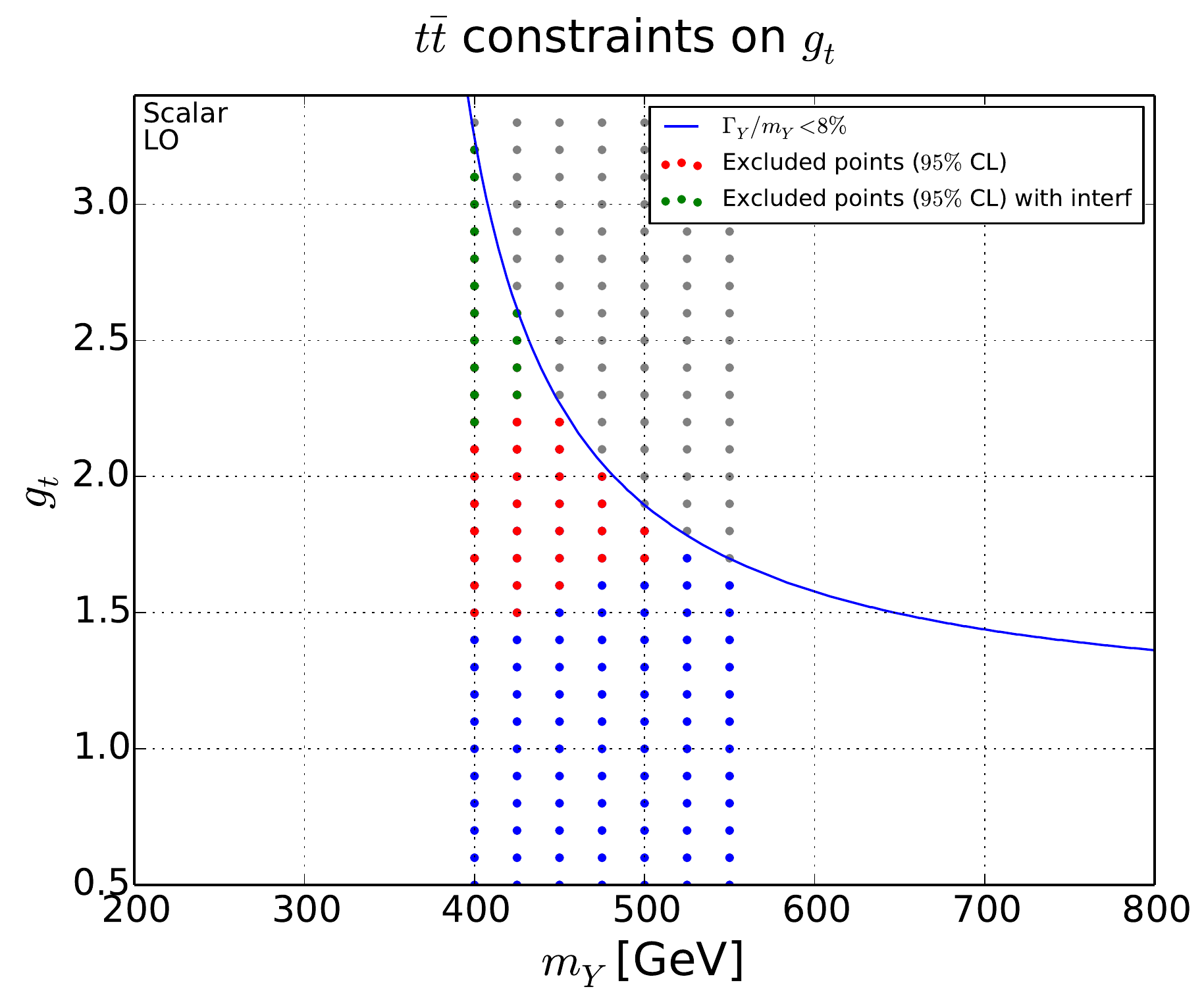}
\includegraphics[scale=0.4]{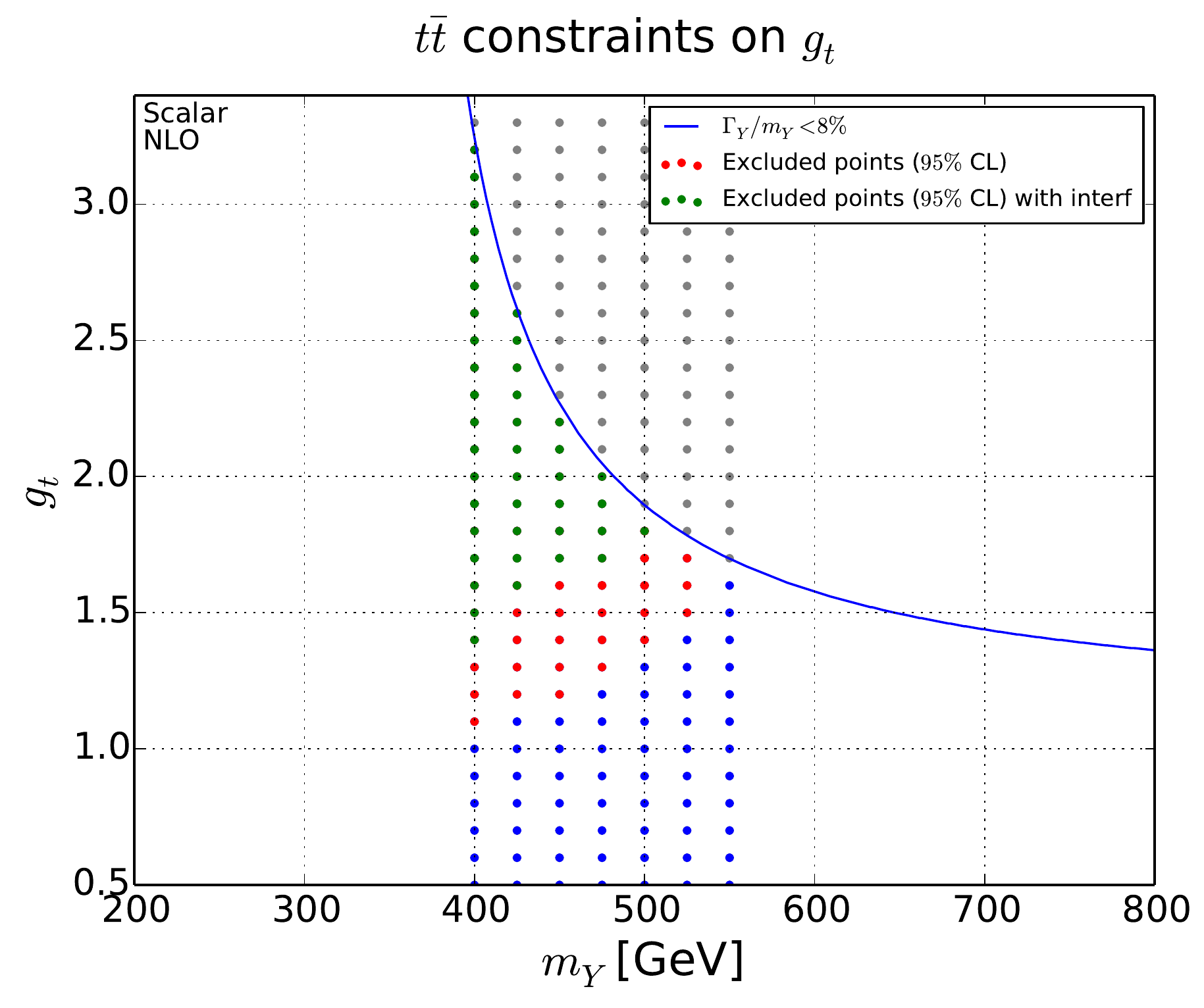}
\caption{Exclusion region obtained using the ATLAS 8~TeV $t\bar{t}$ resonance search results~\cite{Aad:2015fna} for a scalar resonance coupling to the top only using LO (left) and NLO (right) predictions for the BSM cross section. The interference between the signal and background is taken into account. }
\label{fig:exclusioninterf}
\end{figure}

\begin{figure}[h!]
\centering
\includegraphics[scale=0.4]{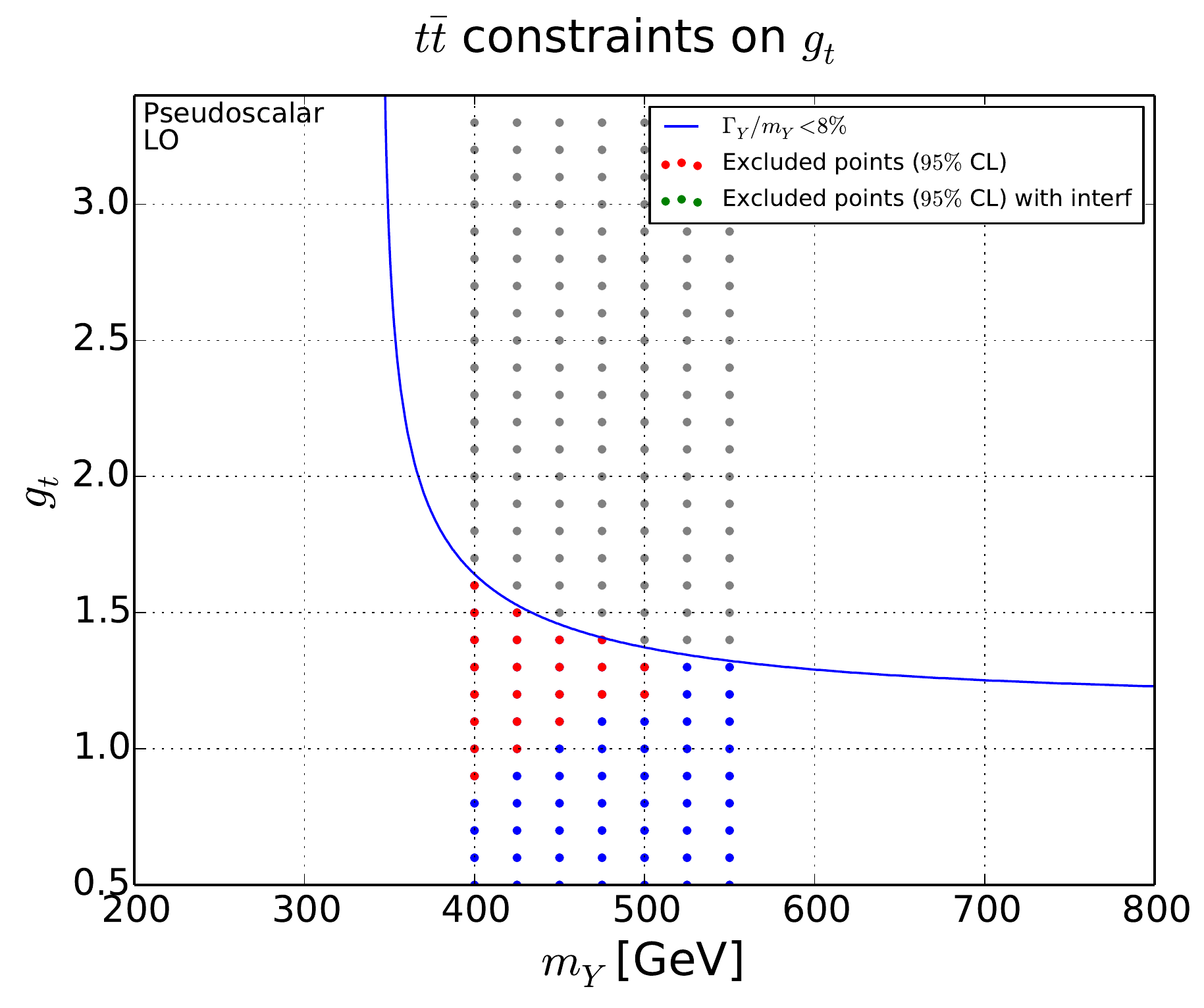}
\includegraphics[scale=0.4]{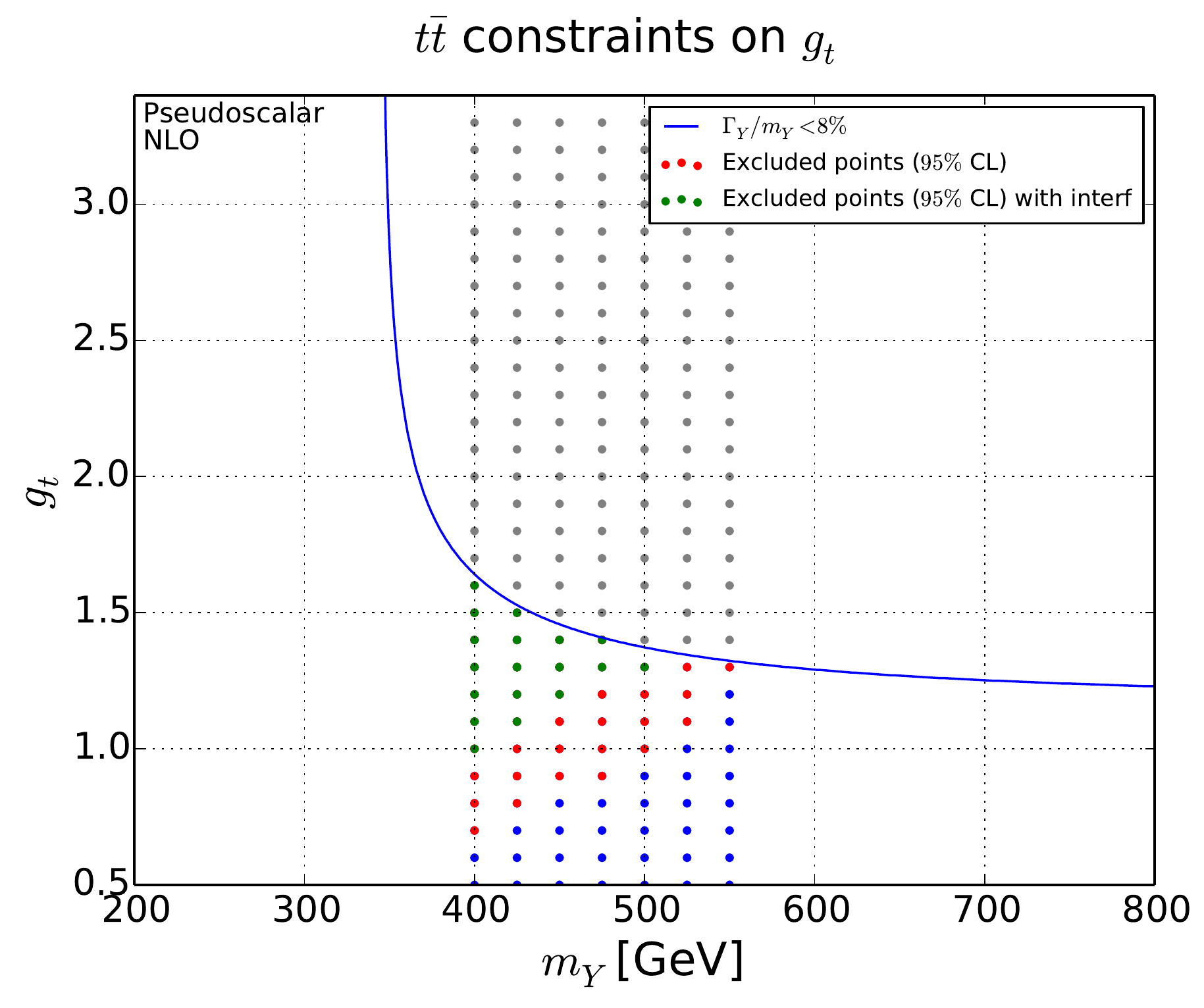}
\caption{Exclusion region obtained using the ATLAS 8~TeV $t\bar{t}$ resonance search results~\cite{Aad:2015fna} for a pseudoscalar resonance coupling to the top only using LO (left) and NLO (right) predictions for the BSM cross section. The interference between the signal and background is taken into account. }
\label{fig:exclusionpseudointerf}
\end{figure}

In order to investigate this, we compute the interference for the parameter points of interest. The results for the exclusion regions are shown in figures \ref{fig:exclusioninterf} and \ref{fig:exclusionpseudointerf}, where the integrated interference rate is simply added to the signal. At LO in the scalar case we see that points with $g_t>2.1$ are excluded even when the interference is taken into account, while for the pseudoscalar no points are excluded which demonstrates the huge impact of the interference in these scenarios. This is particularly evident in the psedoscalar case due to the small coupling restriction imposed by the $8\%$ constraint on the width. The absolute value of the interference can amount up to $50\%$ of the signal in the scalar case and $65\%$ in the pseudoscalar one. At NLO, for both scalar and pseudoscalar resonances, taking into account the interference\footnote{Again we use the LO value multiplied by $\sqrt{K_S K_B}$.} modifies the exclusion region by reducing the number of excluded points. Most of the affected parameter points have cross-sections which were excluded only when computed at NLO. For these points including the destructive interference reduces the cross section enough to fall below the 95\% C.L. limit. 

%While we have studied the impact of the interference only for the simplified model of one additional scalar, similar conclusions can be drawn for the 2HDM and other BSM scenarios. 

\section{750 GeV diphoton excess}
\label{750}
\begin{figure}[h!]
\centering
\includegraphics[scale=0.44,trim=4cm 8cm 0cm 2cm]{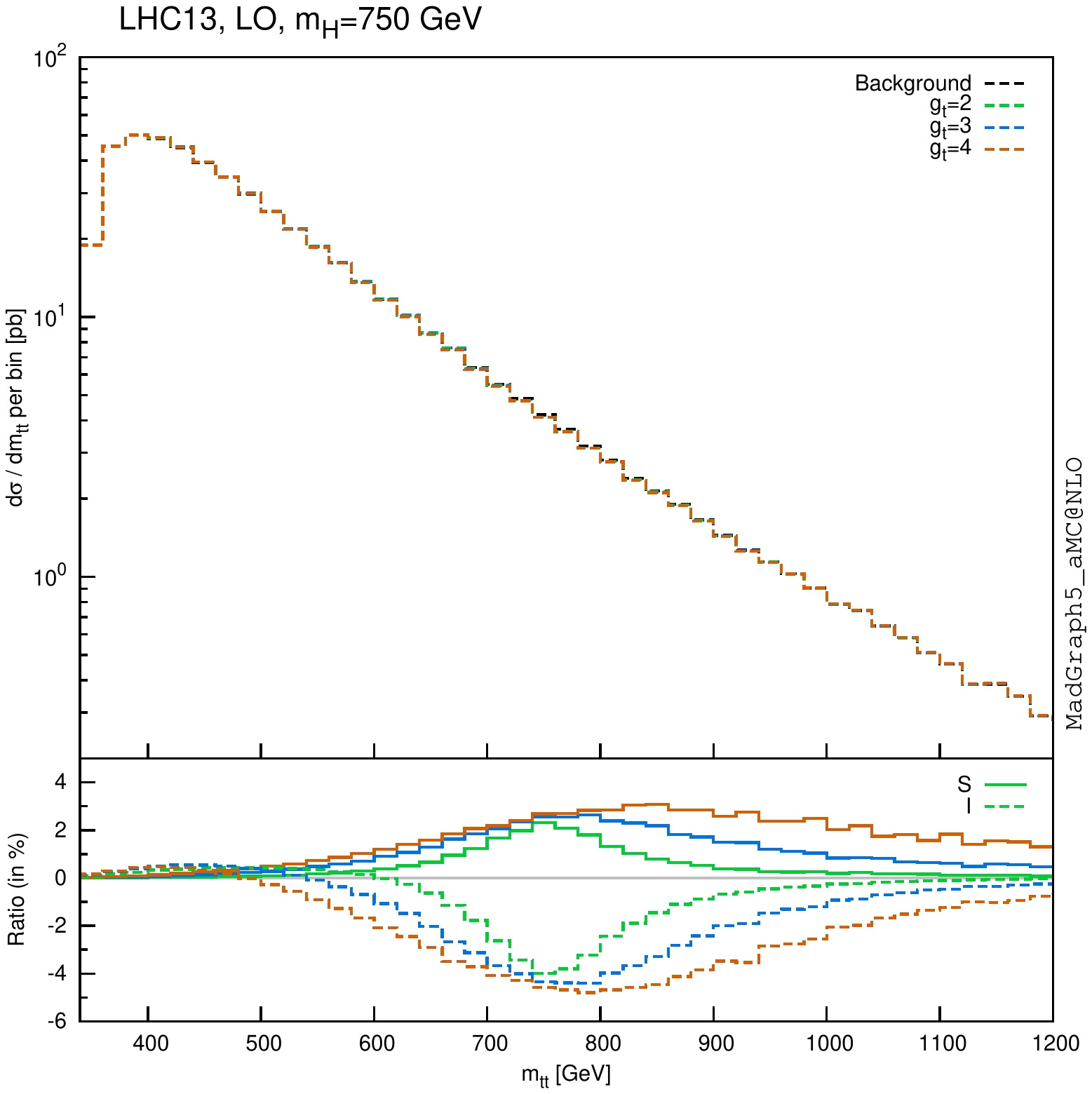}
\includegraphics[scale=0.44,trim=4cm 8cm 4cm 2cm]{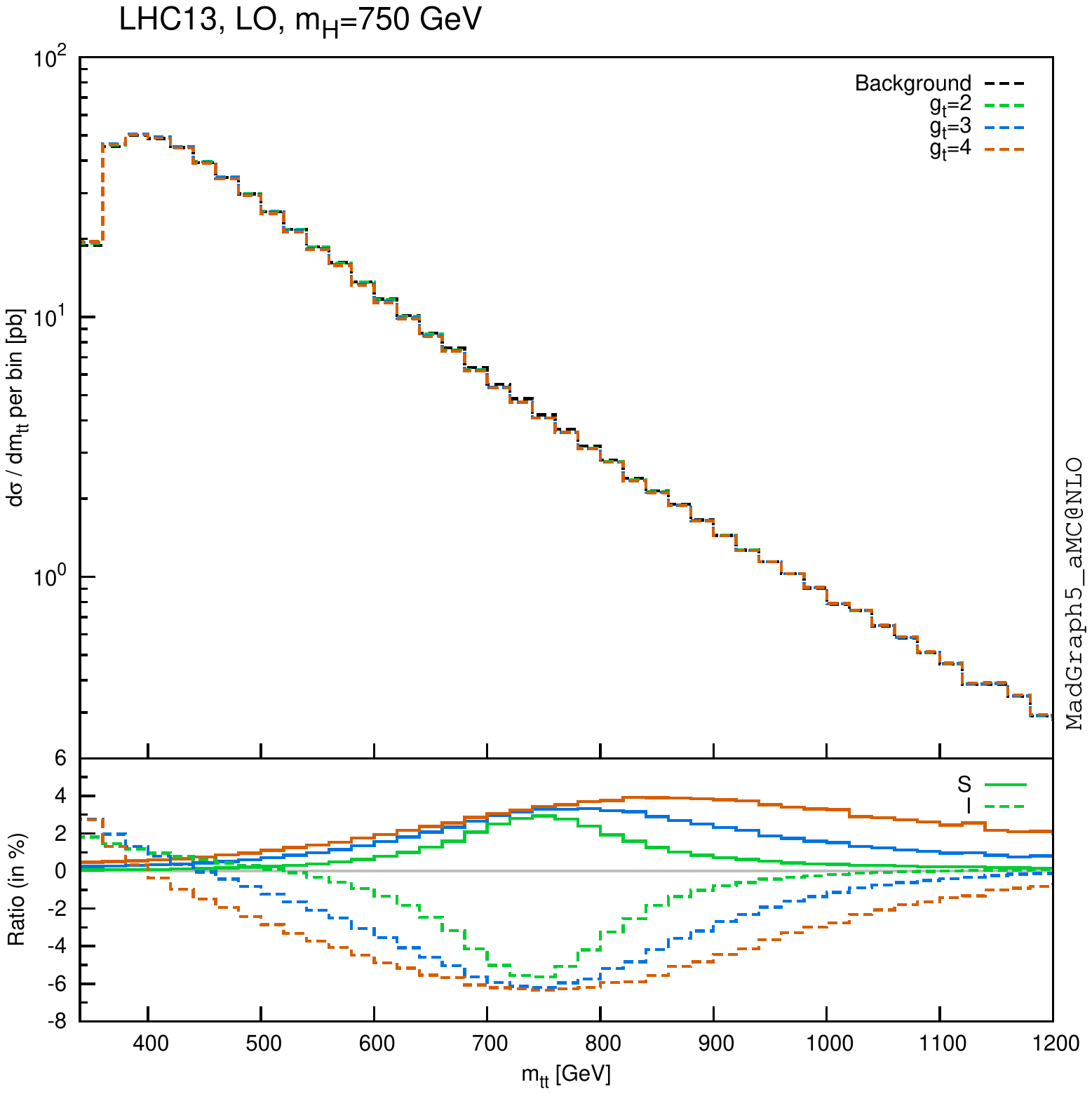}
\caption{Top pair invariant mass distribution at the LHC at 13 TeV for a 750 GeV scalar coupling to the top with different Yukawa couplings. The width of the scalar is computed for each value of the Yukawa. Left: scalar. Right: pseudoscalar. The lower panels show the ratio of signal and interference over the QCD background. }
\label{mtt750}
\end{figure}

In this section we discuss the possible implications of the diphoton excess reported at 750 GeV on top pair production.  The observed excess in
the diphoton spectrum \cite{atlasdiphoton,CMS:2015dxe} is characterised by:
\begin{equation}
m_Y\sim 750\, \textrm{GeV}, \,\, \Gamma_Y/m_Y<6\% \,\, \textrm{and} \,\, \sigma_{\gamma\gamma}\sim 1-10 \textrm{fb}.
\label{diphoton}
\end{equation}
By considering a 750 GeV spin-0 resonance we show the top pair invariant mass distribution in figure \ref{mtt750}. The Yukawa couplings are allowed to vary and the widths are computed accordingly which demonstrates that the resonance becomes very broad for $g_t>1$ and even in that case deviations from the QCD background are at the percent level. Such a model does not give a sufficiently large diphoton signal.  A simple computation shows that a simplified model with a scalar or pseudoscalar resonance coupling only to the top  cannot satisfy the observed features of the excess, as to obtain a sufficiently large production cross section the coupling to the top and consequently the width is forced to be large and beyond perturbative values.

A possible way of enhancing the production cross section without increasing the width beyond the values observed at the LHC is to employ the dimension-5 operators of eq. \ref{gluonoperators}. Even in the presence of these operators we find that in order to satisfy the signal strength properties of the diphoton excess one needs a large coupling to the top to generate the loop suppressed coupling of the scalar to photons. As the dominant decay mode is decay into top-quark pairs, this setup leads to large top-pair cross section values which have already been excluded by the resonant searches. To circumvent this problem one can introduce direct couplings of the scalar to the photons in the following form:
\begin{equation}
L_{\gamma}=-\frac{2\alpha_{EM} c^S_{\gamma}}{9\pi v}F_{\mu\nu}F^{\mu \nu} H^0-\frac{\alpha_{EM} c^P_{\gamma}}{3\pi v}F_{\mu\nu}\tilde{F}^{\mu \nu} A^0, 
\label{photonoperators}
\end{equation}
which are in direct correspondence with those of eq. \ref{gluonoperators} assuming heavy quarks in the loops. 

To investigate the implications of the 750 GeV resonance on top pair production we employ the scalar couplings to tops, gluons and photons. The width of the scalar particle can then be computed from the partial widths to the tops, gluons and photons given by: 
%The partial width to gluons can be computed at tree level in this case, but this is suppressed compared to the one into top. The partial widths are given  by:
\begin{align}
\Gamma(H^0 \to t \bar{t}) = & g_t^2 \frac{3 y_t^2 m_H}{16\pi} \beta_t^3\, \Theta(m_H- 2 m_t)\,, \\
 \displaybreak[1]
\Gamma(A^0 \to t \bar{t}) = & g_t^2 \frac{3 y_t^2 m_A}{16\pi} \beta_t\, \Theta(m_A- 2 m_t)\,, \\
%\end{eqnarray}
%\begin{eqnarray}
\Gamma(H^0  \to g g) = &\frac{\alpha_s^2 m_H^3}{72 \pi^3 v^2} \Big| \frac{3}{2} g_t F_S \Big( \frac{4 m_t^2}{m_H^2} \Big) + c^S_g  \Big|^2\,, \\
 \displaybreak[1]
\Gamma(A^0  \to g g) = &\frac{\alpha_s^2 m_A^3}{32 \pi^3 v^2} \Big| g_t F_P \Big( \frac{4 m_t^2}{m_A^2} \Big) + c^P_g  \Big|^2\,, \\
%\end{eqnarray}
%\begin{eqnarray}
 \Gamma(H^0  \to \gamma \gamma ) =& \frac{\alpha_e^2 m_H^3}{81 \pi^3 v^2} \Big| \frac{3}{2} g_t F_S \Big( \frac{4 m_t^2}{m_H^2} \Big) + c^S_{\gamma}  \Big|^2\,,   \\
 \displaybreak[1]
 \Gamma(A^0 \to \gamma \gamma ) =& \frac{\alpha_e^2 m_A^3}{36 \pi^3 v^2} \Big|g_t F_P \Big( \frac{4 m_t^2}{m_A^2} \Big) + c^P_{\gamma}  \Big|^2\,,    \label{eq:y0widthph}
\end{align} with 
\begin{eqnarray}
\label{eq:fsg}
\beta_t &=& \sqrt{1-\frac{4m_t^2}{m_{H/A}^2}}, \\
F_{S}(x) &=& x\Big[1 + (1-x) \, {\rm \arctan}^2 \Big( \frac{1}{\sqrt{x-1}} \Big)\Big]\,, \\
F_{P}(x) &=& x\, {\rm \arctan}^2 \Big( \frac{1}{\sqrt{x-1}} \Big)\,.
\end{eqnarray}

\begin{figure}[h!]
\centering
\includegraphics[scale=0.55,trim=10cm 0cm 0cm 0cm]{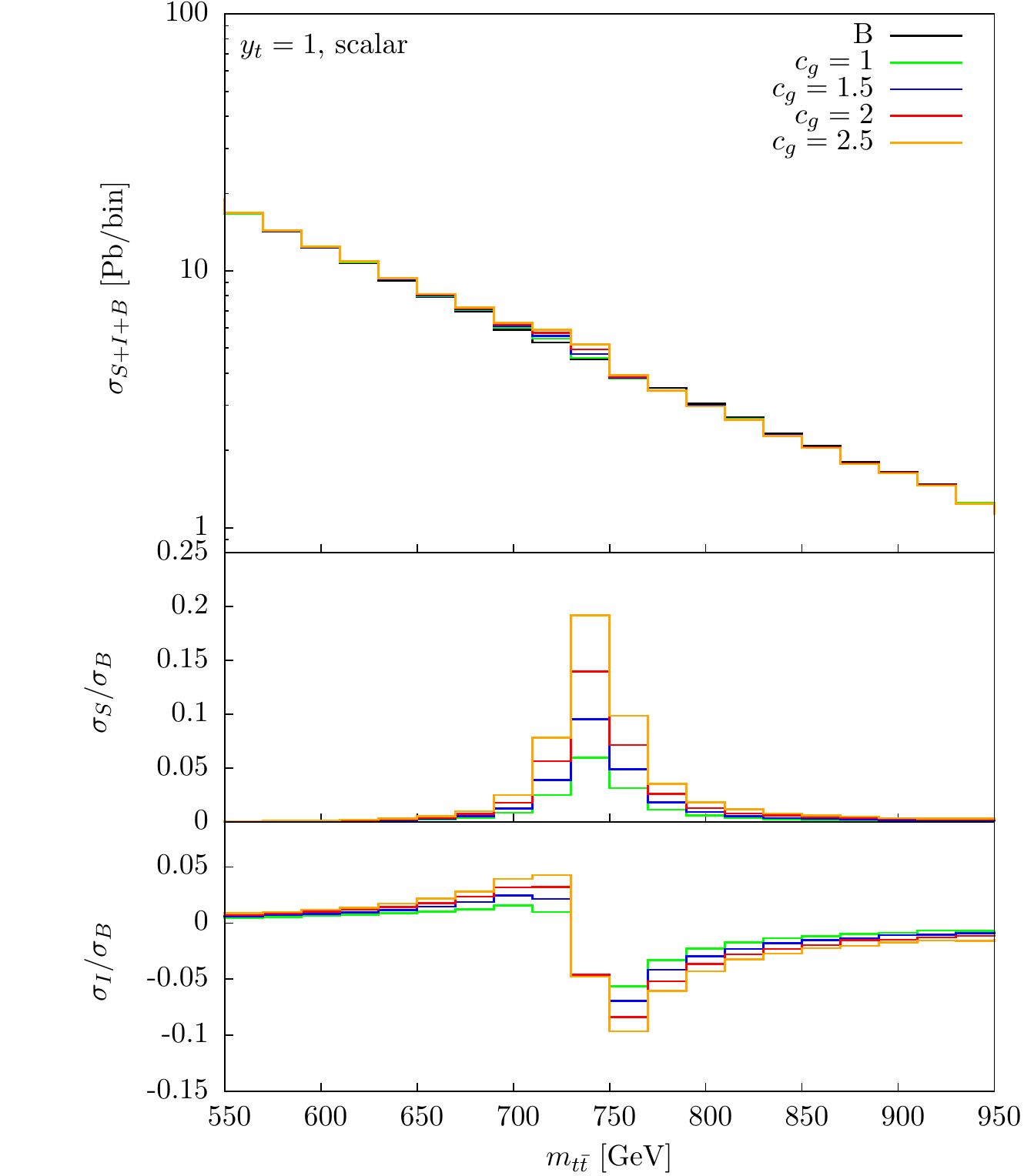}
\includegraphics[scale=0.55,trim=1cm 0cm 8cm 0cm]{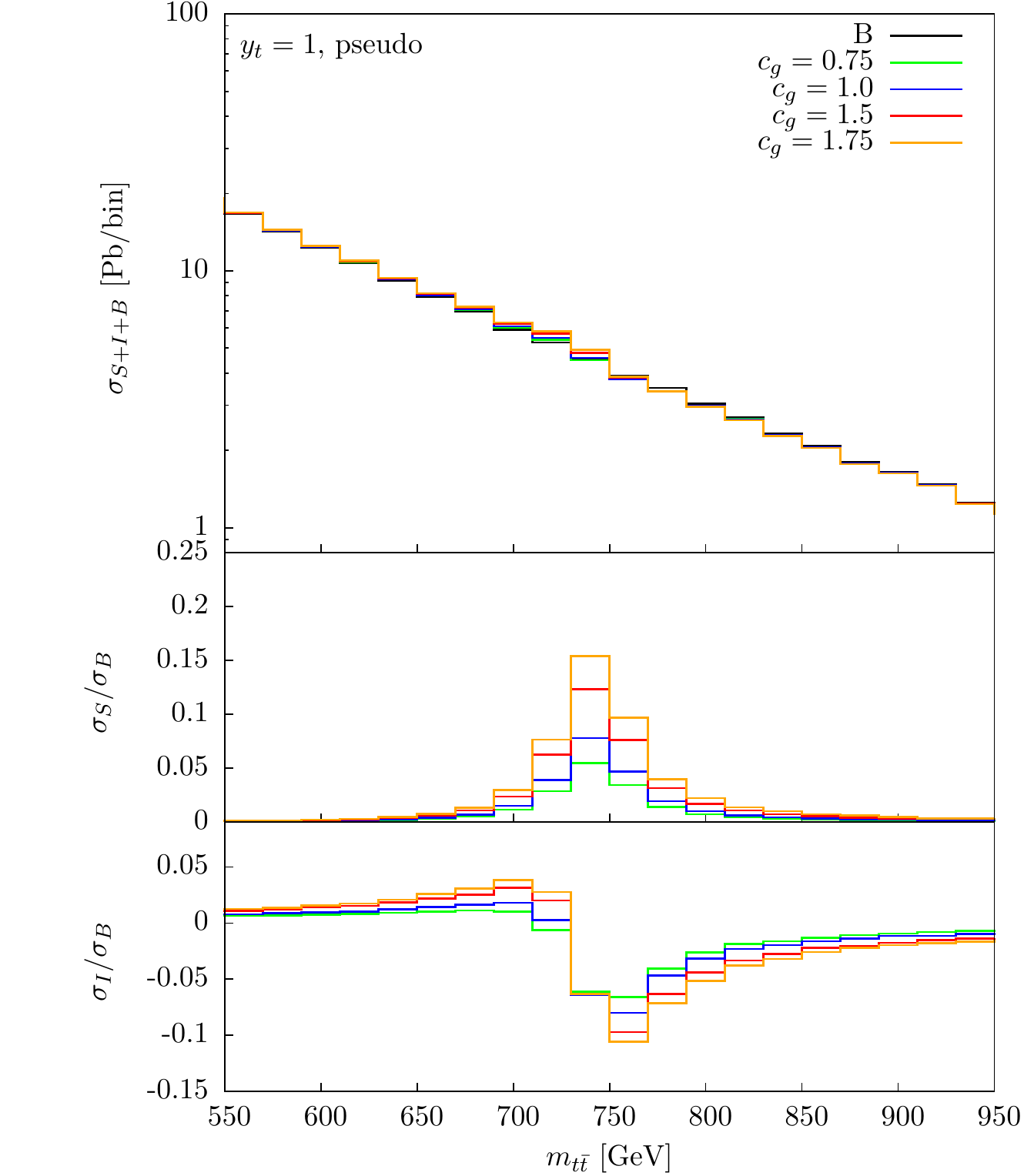}
\caption{Top pair invariant mass distribution for the LHC at 13 TeV in the presence of a 750 GeV resonance coupling to gluons, photons and top quarks. The values of the couplings shown here satisfy the diphoton excess properties. The lower panels show the ratio of the signal and interference over the QCD background. }
\label{mtt750_ggy}
\end{figure}

A selection of possible parameter setups which satisfy the diphoton observations of eq.~\ref{diphoton} is shown in table \ref{table:effparam} along with the scalar width, the diphoton and top--anti-top signal cross-sections computed in the narrow width approximation using NLO cross sections for the scalar production. We note that we have checked explicitly that the $t\bar{t}$ cross-section is smaller than what one would exclude at 750 GeV using the LHC resonant search results, despite the fact that the top branching ratio exceeds 95\% for all scenarios listed here. For this selection of benchmark points, we present results for the signal and signal-background interference in top pair production in figure \ref{mtt750_ggy} for a scalar or pseudoscalar resonance of 750 GeV. In all cases the interference should be taken into account and has a significant impact on the line-shape of the resonance.   

\begin{table}[h!]
\begin{center}
    \begin{tabular}{|c|c|c|c|c|c|c|}
        \hline 
&$g_t$   & $c_g$  & $c_{\gamma}$ & $\Gamma_{tot}$ & $\sigma(pp \to Y \to \gamma \gamma)$ & $\sigma(pp \to Y \to t \bar{t})$ \\ \hline

\multirow{4}{*}{Scalar} &1 & 1.0 & 100 & 32.8 & 9.4 fb & 0.2 pb \\ \cline{2-7} 
&1 & 1.5 & 55 & 31.7 & 6.7 fb & 0.4 pb \\ \cline{2-7} 
&1 & 2.0 & 30 & 31.4 & 3.6 fb & 0.7 pb \\ \cline{2-7} 
&1 & 2.5 & 20 & 31.4 & 2.5 fb & 1.1 pb \\ \hline \hline
\multirow{4}{*}{Pseudoscalar}&1 & 0.75 & 65 & 41.1 & 9.0 fb & 0.2 pb \\ \cline{2-7} 
&1 & 1.0 & 45 & 40.3 & 7.8 fb & 0.4 pb \\ \cline{2-7}
&1 & 1.5 & 20 & 39.8 & 3.6 fb & 0.9 pb \\ \cline{2-7} 
&1 & 1.75 & 10 & 39.7 & 1.2 fb & 1.2 pb \\ \hline 
\end{tabular}
 \caption{Example of benchmarks points in our simplified model satisfying the currently available information on the diphoton excess. The couplings of the scalar to tops, gluons and photons are given along with the scalar width and the narrow width diphoton and $t\bar{t}$ signal cross sections for a 750 GeV scalar or pseudoscalar resonance.  \label{table:effparam}}
\end{center} 
\end{table}

\section{Conclusions}
\label{conclusions}

We have studied the interference between a new physics signal and the QCD background in the presence of additional scalars that decay into top quark pairs. The interference with the background needs to be taken into account to reliably predict the line-shape of the additional scalar. We have explored the impact of the interference within a simplified model with a heavy scalar, pseudoscalar or mixed state as well as for a set of representative 2HDM scenarios. The interference leads to interesting peak-dip features in the invariant mass distribution of the top pair. While the observed features depend on the specific model parameters, we find that in general the impact of the interference becomes rapidly important once the width over mass ratio of the resonance rises above a few percent.

In addition to the $t\bar{t}$ process, the interference has been studied when the top pair is produced in association with a jet. We find that the size and shape of the interference compared to the background is not significantly modified compared to the $2\to 2$ process but remains important in the determination of the shape of the invariant mass distribution. In order to improve the precision for the signal process we have computed it at NLO accuracy in QCD. We find large QCD corrections for all scenarios studied. While an exact NLO computation for the interference is beyond recent advances in loop technology, we approximate the interference at NLO using the geometric average of the signal and background $K$-factors, which provides an estimate of the higher order QCD effects. This procedure has also been validated by the $t\bar{t}+$jet calculation. 

For a simplified model of an additional scalar coupling to the top only, we have studied the region of the parameter space of the model that can be excluded by the ATLAS top pair narrow-width resonance search. This simple scenario demonstrates the importance of taking into account both the NLO corrections and the interference with the QCD background when setting limits on BSM scenarios. While in our analysis only total rates have been used to set limits on the parameter space of the model, it is important to stress that the shapes of the distributions are significantly changed by the interference and the experimental analyses should be accordingly modified to account for this, in particular as they extend their search beyond the narrow width approximation. 

Finally we have also discussed the implications of the recently reported 750 GeV diphoton excess on top pair production. We have explored a scenario with a 750 GeV scalar boson coupling to gluons and photons through an effective coupling and in addition directly to top quarks. For parameters satisfying the characteristics of the excess we find that again the interference with the QCD background needs to be taken into account when searching for signs of the resonance in the top--anti-top channel. 

\acknowledgments
We would like to thank David L\'opez-Val for his assistance with the 2HDM benchmarks and Marius Wiesemann for his assistance with aMCSusHi.  This work has been performed in the framework of the ERC Grant No. 291377 ``LHCTheory''
and has been supported in part by the European Union as part of the FP7 Marie Curie Initial Training Network MCnetITN  (PITN-GA-2012-315877) and by the National Fund for Scientific Research (F.R.S.-FNRS Belgium) under a FRIA grant.

\bibliographystyle{JHEP}
\bibliography{zhbib}
\end{document}